\documentclass[pre,twocolumn,superscriptaddress]{revtex4}
\usepackage{amsmath}
\usepackage{amssymb}

\usepackage{amsmath,amssymb}
\usepackage[usenames]{color}
\usepackage{amssymb}
\usepackage{grffile}
\usepackage[pdftex]{graphicx}
\usepackage{amsmath, amstext, amssymb, amsfonts, amsxtra}
\usepackage{textcomp}
\usepackage{xspace} 
\usepackage[colorlinks]{hyperref}

\DeclareSymbolFontAlphabet{\amsmathbb}{AMSb}
\newcommand{\sv}{\vec{\sigma}}
\newcommand{\s}{\sigma}
\newcommand{\tav}{\vec{\tau}}
\newcommand{\ta}{\tau}
\newcommand{\km}{k_{max}}
\newcommand{\al}{\alpha} 
\newcommand{\ep}{\epsilon}    
\newcommand{\err}{\varepsilon}

\newcommand{\tr}{{\rm tr}}

\usepackage{tikz}
\usepackage{pgf}
\usepackage{tikz-qtree}
\usetikzlibrary{arrows,decorations.pathmorphing,backgrounds,positioning,fit,petri,shapes.misc, arrows.meta,shapes.geometric,decorations.markings,calc,shadows.blur}
\definecolor{myblue}{RGB}{20,162,212}
\definecolor{myorange}{RGB}{211, 84, 0}
\definecolor{lowblue}{RGB}{102,178,255}
\definecolor{mypurple}{RGB}{142, 68, 173}
\definecolor{mygrey}{RGB}{158, 158, 158}
\definecolor{lowpurple}{RGB}{204,153,255}
\definecolor{lowwhite}{RGB}{255,255,255}
\definecolor{verylowpurple}{RGB}{255,102,102}
\definecolor{embcolor}{RGB}{255,255,255}
\definecolor{myred}{RGB}{231, 76, 60}
\definecolor{mygreen}{RGB}{162, 217, 206} 
\definecolor{fontgrey}{RGB}{44, 62, 80}
\definecolor{lowpurple}{RGB}{210, 180, 222}
\definecolor{mypumpkin}{RGB}{229, 152, 102}
\definecolor{lowgreen}{RGB}{171, 235, 198}
\definecolor{lowred}{RGB}{245, 183, 177}

\newcommand{\epd}{Engineering Product Development Pillar, Singapore University of Technology and Design, 8 Somapah Road, 487372 Singapore}
\newcommand{\istd}{Information Systems Technology and Design, Singapore University of Technology and Design, 8 Somapah Road, 487372 Singapore}
\newcommand{\zz}{Zhengzhou Information Science and Technology Institute, Zhengzhou 450004, China}

\begin{document}

\title{Matrix Product Operators for Sequence to Sequence Learning}

\author{Chu Guo}
\affiliation{\epd} 
\affiliation{\zz} 
\author{Zhanming Jie} 
\affiliation{\istd}    
\author{Wei Lu}
\affiliation{\istd} 
\author{Dario Poletti}
\affiliation{\epd} 

\begin{abstract}
The method of choice to study one-dimensional strongly interacting many body quantum systems is based on matrix product states and operators. Such method allows to explore the most relevant, and numerically manageable, portion of an exponentially large space. It also allows to describe accurately correlations between distant parts of a system, an important ingredient to account for the context in machine learning tasks. Here we introduce a machine learning model in which matrix product operators are trained to implement sequence to sequence prediction, i.e. given a sequence at a time step, it allows one to predict the next sequence. We then apply our algorithm to cellular automata (for which we show exact analytical solutions in terms of matrix product operators), and to nonlinear coupled maps. We show advantages of the proposed algorithm when compared to conditional random fields and bidirectional long short-term memory neural network. To highlight the flexibility of the algorithm, we also show that it can readily perform classification tasks.       	 	 	        
\end{abstract}

\date{\today}
\maketitle


The last few years have witnessed a great shift of interest of society (individuals, industries and governmental organizations) towards machine learning and its wide range of applications ranging from images classification to translation and more.     

In recent years we have also witnessed an increasing activity at the intersection between machine learning and quantum physics. This includes further studies on quantum machine learning \cite{SchuldPetruccione2014, AdcockStanisic2015, BiamonteLloyd2016}, use of machine learning for materials study \cite{KalininArchibald2015}, glassy physics \cite{SchoenholzLiu2016}, for phases recognition \cite{Wang2016, CarrasquillaMelko2016, ChngKhatami2016, BroeckerTrebst2016, NieuwenburgHuber2017, ZhangKim2017, ChngKhatami2018, ZhangZhai2018} and to numerically study quantum systems \cite{ArsenaultMillis2014, ArsenaultMillis2015, CarleoTroyer2017, AminMelko2016, AokiKobayashi2016, HuangWang2016, LiuPu2016, TorlaMelko2016, NomuraImada2017, CzischekGasenzer2018}. To be noted are studies on physical analogies and reasons for effectiveness of machine learning \cite{Beny2013, MethaSchwab2014, LinRolnick2017}. 

Another research direction has been to apply tools developed in many body quantum physics to typical machine learning tasks. Recent examples are \cite{StoudenmireSchwab2016, HanZhang2017, Stoudenmire2017}, where algorithms based on matrix product states (MPS), also known as tensor trains, were successfully used for supervised classification or unsupervised generative modelling.         
MPS based algorithms were also used for classification \cite{NovikovOseledets2017}, as predictive modeling of stochastic processes \cite{YangGu2018}, for language modeling \cite{PestunVlassopoulos2017} and compared to Boltzmann machines \cite{DengSarma2017, ChenXiang2018}. Usefuleness of tensor representations and matrix product states was also noted in \cite{Cichocki2014}. 
 
In physics, MPS are used to represent wave-functions, probability distributions or density matrices as a product of tensors \cite{DerridaPasquier1993, KrebsSandov1997, JohnsonJaksch2010, Schollwock2011, JohnsonJaksch2015}. Building on the density matrix renormalization group \cite{White1992}, matrix product states are very succesfully used in many body quantum physics to study ground or steady states and time evolution of Hamiltonian or dissipative systems \cite{WhiteFeguin2004, Vidal2004, DaleyVidal2004, VerstraeteCirac2004, Schollwock2005, McCulloch2007, Daley2014}. As an extension, quantum mechanical operators are represented by matrix product operators (MPO), which are composed of tensors and map an MPS to another one. One important aspect of matrix product states is that they can allow to give an accurate description of correlations between distant parts of a system. This feature can be of great use for machine learning models in which the algorithm needs to be able to consider accurately the context. 

In this work we implement a matrix product operator approach for sequence to sequence prediction. Based on a set of input and output vectors, we are going to train an MPO which can accurately reproduce the transformation from input to output vectors and which can be then used for vectors not used in the training. As exemplary applications we are going to train an MPO to predict the evolution of dynamical systems, in particular cellular automata and coupled maps. 
We will show that it is possible to train an MPO to predict exactly the evolution of cellular automata, as long as the matrix product operators are large enough. It is even possible to write analytically an MPO description for cellular automata which we discuss both in the main text and in the appendix. We will then show that the predictions of the trained MPO for the evolution of cellular automata can be perfect even in presence of a large portion of errors in the training data.  
We compare our MPO-based method to conditional random fields, CRF \cite{LaffertyPereira2001}, and discuss the respective advantages of the two methods. 
Later we apply our algorithm to the prediction of the evolution of a nonlinear coupled map. Despite the lack of exact analytical solutions for the MPO, we show that it can be used to predict the evolution of such systems as well. When compared to a bidirectional long short-term memory (LSTM) neural network \cite{HochreiterSchmidhuber1997}, which is a widely-used sequence prediction machine learning model, we show that the MPO algorithm gives accurate and comparable results for the problem analyzed.    

To exemplify the generality of the algorithm, we also describe how to apply it to classification problems, using as an example the MNIST handwritten digits data set.  


\section{Method}\label{sec:model}
\begin{figure}
\includegraphics[width=\columnwidth]{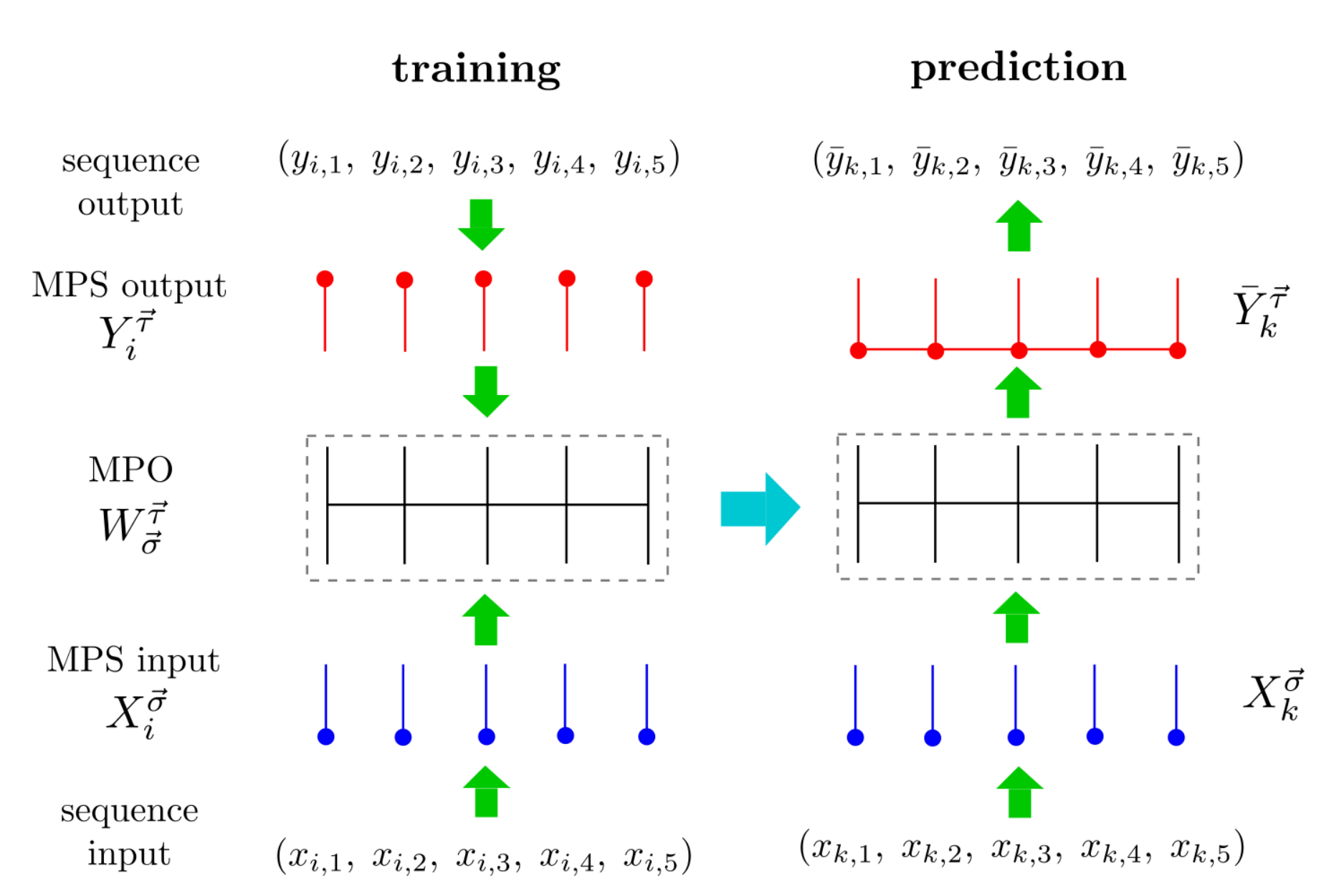}
\caption{Training phase: a pair of vectors, $\vec{x}_i$ and $\vec{y}_i$, is converted to a pair of matrix product states (MPS), $X^{\vec{\sigma}}_i$ and $Y^{\vec{\tau}}_i$, and used to train a matrix product operator (MPO) $W^{\tav}_{\sv}$. Prediction phase: an input vector $\vec{x}_k$ is converted to an MPS $X^{\vec{\sigma}}_k$ and then multiplied by the trained MPO $W^{\tav}_{\sv}$. The resulting MPS $\bar{Y}^{\vec{\tau}}_i$ is then converted to a vector $\vec{\bar{y}}_k$.   
  \label{fig:Fig1} }
\end{figure}

In the following section we describe how we implement the training (or learning phase) of our setup and how to test the accuracy of its predictions (testing phase). The MPO algorithm will be able to generate a sequence of $L$ numbers given another one of same length. 
Stated differently, we consider mapping from a vector space $\mathcal{X}$, with vectors $\vec{x}_i$, to another vector space $\mathcal{Y}$, with vectors $\vec{y}_i$, which is given by 
\begin{align}
\vec{y}_i = f(\vec{x}_i),
\end{align}
where $f$ does not need to be linear. The MPO algorithm would try to provide an accurate approximation $\vec{\bar{y}}_i$ to the exact vector $\vec{y}_i$.

We will focus on the case for which each vector $\vec{x}_i, \vec{y}_i$ have the same length $L$ and corresponding elements $x_{i,l}$ and $y_{i,l}$. The algorithm will be explained in detail in the following, however, to help the reader we summarize it in Fig.(\ref{fig:Fig1}). In the training phase we take a set of $M$ input and output sequences $\vec{x}_i$ and $\vec{y}_i$, where $1 \leq i \leq M$. Each sequence $\vec{x}_i$ and $\vec{y}_i$ is converted to a matrix product state (MPS) -- that is a product of tensors. All these MPS are used to generate/train a mapping from MPS to MPS which is known as matrix product operator (MPO). 

In contrast, in the testing/prediction phase, see right portion of Fig.\ref{fig:Fig1}, we take an input sequence $(x_{k,1},x_{k,2},\dots,x_{k,L})$ (different from those used in the training), convert it to an MPS, multiply it to the MPO trained earlier and this will result in an output MPS which is then converted to a sequence $(\bar{y}_{k,1},\bar{y}_{k,2},\dots,\bar{y}_{k,L})$. To compute the accuracy of the algorithm, we compare the predicted output sequence with the exact one.

\subsection{Training} \label{ssec:training}     

The algorithm aims to provide, given a sequence, another sequence from an exponentially large number of possible one. The output will thus be a function of the possible output sequences, and it will be used to choose the output sequence. One key ingredient in the proposed algorithm is to describe inputs and outputs, or their probability distribution, as MPS \cite{DerridaPasquier1993, KrebsSandov1997, JohnsonJaksch2010, JohnsonJaksch2015}. For example, the probability of a given sequence of integer numbers $\vec{\sigma}=(\sigma_1,\sigma_2,\dots,\sigma_L)$, i.e. $P(\vec{\sigma})$, can be written as     
\begin{align}
P(\vec{\sigma}) = \sum_{a_0, \dots, a_L} M_{a_0, a_1}^{\sigma_1}M_{a_1, a_2}^{\sigma_2} \dots M_{a_{L-1}, a_L}^{\sigma_L},
\end{align}
where $\sigma_l$ describe the local degrees of freedom, whose total number is $d$, and the $a_l$ are auxiliary degrees of freedom needed to take into account correlations between different elements of the sequence. The larger the number of the auxiliary local degrees of freedom $a_l$, also known as ``bond dimension'' $D$, the more accurate is the probability distribution. A single sequence $\vec{\sigma}$ can be written as an MPS with $D=1$, also known as a product state.   
Since we rewrite inputs and outputs as MPS the natural object to train, which will then be able to map an input MPS to an output MPS, is an MPO $W_{\sv}^{\tav}$ where $\tav$ is a sequence of output integer numbers. 
The MPOs can be parameterized by a product of $4$-dimensional tensors 
\begin{align}
W_{\sv}^{\tav} = \sum_{b_0, b_1, \dots, b_L} W_{b_0, b_1}^{\sigma_1, \ta_1}W_{b_1, b_2}^{\sigma_2, \ta_2} \dots W_{b_{L-1}, b_L}^{\sigma_L, \ta_L}
\end{align}
Here, the indices $b_l$ indicate an auxiliary dimension which we refer to as the MPO's bond dimension $D_W$. 

In the next sections we are going to describe how to convert a sequence (also of real numbers) to an MPS and then how to train an MPO. 

\subsubsection{From sequence to MPS}
The first necessary step to train an MPO is to convert a sequence (input or output) to an MPS. The input and output sequences may belong to different spaces, for example an input sequence could be made of words, letters, real numbers or integers. Each of the above can be represented as a vector of finite size $d$ for the input, and $d'$ for the output (which can potentially be different). Each sequence is thus readily mapped to a list of vectors.   
For clarity of explanation, in the following we will consider the cases in which $x_{i,l}$ is an integer number or a real number between $0$ and $1$. A completely analogous discussion can be done for the output sequences with elements $y_{i,l}$. 
If $x_{i,l}$ is an integer number which spans over $d$ possible values, it is then mapped to a vector with all $0$ entries except for one of them,  corresponding to a particular $\s_l$, which is set equal to $1$ (this representation is known as one hot vector). This means that considering $x_{i,l}=0,\;1$ or $2$, then $d=3$ and, for example, if $x_{i,l}=1$ we get $\vec{\sigma}_l=(0,1,0)$. 
If $x_{i,l}$ is instead a real number between $0$ and $1$, it can be mapped, for example, to a vector with dimension $2$ with elements $\left(\sqrt{1-x_{i,l}^2}\;, x_{i,l} \right)$.

In this way, each input sequence $\vec{x}_i$ can be mapped into a product MPS of physical dimension $d$.
\begin{align}
\vec{x}_i \rightarrow X_i^{\sv} &= \!\!\sum_{a_0, \dots, a_L}\! X_{i, a_0, a_1}^{\sigma_1}X_{i, a_1, a_2}^{\sigma_2} \dots X_{i, a_{L-1}, a_L}^{\sigma_L} \label{eq:vector} 
\end{align}
where all the auxiliary indices $a_l=1$ (bond dimension $D=1$), and the $i$ index in the tensors $X^{\sigma_l}_{i,a_l,a_{l+1}}$ differentiate the $i-$th sequence from others. The index $\sigma_l$ signals which element of the $d$ dimensional vector is being considered. As an example, for the case in which $x_{i,l}$ is a real number between $0$ and $1$, each matrix product state $X^{\sigma_l}_{i,a_l,a_{l+1}}$ is given by 
\begin{align}
X^{0}_{i,1,1} = \sqrt{1-x_{i,l}^2} \;\; {\rm and} \;\;  X^{1}_{i,1,1} =  x_{i,l}.
\end{align} 
Note that a similar mapping was done in \cite{StoudenmireSchwab2016}.    

Analogously, each output $\vec{y}_i$ can be mapped into a tensor product MPS
\begin{align}
\vec{y}_i \rightarrow Y_i^{\tav} &= \!\!\sum_{c_0, \dots, c_L} \! Y_{i, c_0, c_1}^{\tau_1}Y_{i, c_1, c_2}^{\tau_2} \dots Y_{i, c_{L-1}, c_L}^{\tau_L} 
\end{align}
where the auxiliary indices $c_l=1$, while the index $\ta_l$ signals which element of the $d'$ dimensional local output vector is being considered.

As discussed earlier, multiplying an MPS by an MPO results in another MPS 
\begin{align}
\bar{Y}_i^{\tav} &= W_{\sv}^{\tav} X_i^{\sv} = \sum_{\bar{c}_0, \dots, \bar{c}_L}\bar{Y}_{i, \bar{c}_0, \bar{c}_1}^{\ta_1} \dots \bar{Y}_{i, \bar{c}_{L-1}, \bar{c}_L}^{\ta_L},
\end{align}
where $\bar{c}_l$ is a new auxiliary index given by $\bar{c}_l=(a_l,b_l)$ on each site $l$, and where we have used
\begin{align}
\bar{Y}_{i, \bar{c}_{l-1}, \bar{c}_l}^{\ta_l} = \sum_{\s_l}W_{b_{l-1}, b_l}^{\s_l, \ta_l} X_{i,a_{l-1}, a_l}^{\s_l}.
\end{align}

\subsubsection{Iterative minimization of the cost function}     
With the above notations, we can define a cost function $C(W_{\sv}^{\tav})$, which takes into account the distance between all the predicted and the exact output sequences. The cost function we use is   
\begin{align}\label{eq:quantum_cost}
C(W_{\sv}^{\tav}) &= \sum_{i=1}^N \left(\bar{Y}_i^{\tav\dagger} - Y_i^{\tav\dagger}  \right)\left( \bar{Y}_i^{\tav} - Y_i^{\tav}  \right) \nonumber\\ 
&+\alpha\; \tr\left(W_{\sv}^{\tav\dagger} W_{\sv}^{\tav}\right),                         
\end{align}
where the last term with the coefficient $\alpha$ regularizes the MPO. 

To minimize the cost function, we use an iterative approach, similar to the MPS method used for ground state search for one-dimensional quantum many body problems \cite{Schollwock2011}. The central idea is to turn a global minimization problem into many local minimization problems via an iterative procedure. To minimize the cost function we compute its derivative versus the local tensor $W_{b_{l-1}, b_l}^{\sigma_l, \ta_l}$ of $W_{\sv}^{\tav}$, and set it to zero    
\begin{align}
\frac{\partial C(\hat{W})}{\partial W^{\sigma_l, \ta_l}_{b_{l-1}, b_l}} = 0. \label{eq:minimize}      
\end{align}
The minimization procedure is done from site $l=1$ to site $l=L$ and back, and this procedure is commonly referred to, in many body quantum physics, as sweep. Since the cost function is quadratic, a linear solver can be used to compute the locally optimal tensor $W_{b_{l-1}, b_l}^{\sigma_l, \ta_l}$.  
The sweeps are repeated until a maximum number $\km$ or until the cost function converges to a previously determined precision $\epsilon_t$. The algorithm can be run with different bond dimensions $D_W$ (the size of the auxiliary space) and with different training samples number $M$ in order to reach more accurate sequence predictions.

\subsection{Predictions}     
Once the MPO has been trained, it is possible to use it to make predictions. In order to do so, first the input sequence is converted into an input MPS. This MPS is then multiplied with the trained MPO, and this results in the output matrix product state $\bar{Y}_i^{\tav}$, see also Fig.\ref{fig:Fig1}. It is then necessary to convert the output MPS into an output sequence, and this is done in two steps. First we approximate the output MPS with an MPS of bond dimension $D=1$ (this can done iteratively by minimizing the distance between the output MPS and the approximated one \cite{Schollwock2011}). 
Then, the MPS with bond dimension $D=1$ is converted to a sequence by reversing the way in which, in precedence, a sequence $\vec{x}_i$ (or $\vec{y}_i$) was converted into a $D=1$ matrix product state $X_i^{\sv}$ (or $Y_i^{\tav}$). For example, when $x_{i,l}$ is a real number between $0$ and $1$, then we set $x_{i,l}=X^1_{i,1,1}$; when $x_{i,l}$ is integer, we set $x_{i,l}=\sigma_l$ for the $\sigma_l$ for which $X_{i,1,1}^{\sigma_l}$ is largest.

\section{Applications}\label{sec:applications} 
We are now going study how this algorithm can be used to predict the evolution of different systems. In general one can consider systems which are extended in space, evolve in time, and take different values at each different location and time. We will first concentrate on {\it cellular automata}, for which time, space and the possible values are all discrete. We will then consider {\it coupled maps}, for which, while time and space are discretized, the possible values that a function has at a certain time and location are continuous. 

\subsection{Cellular automata}

Cellular automata are models in which time and space are discretized and when at each location and time, a function takes discrete values. The values of this function at later times are determined by the values at previous time in a neighbourhood of the site considered. Such models have been used in many fields from mathematics to physics, including computer science and biology \cite{Wolfram1983}. 
In the models that we study here, we consider sequences of $0$s and $1$s. At first, to each input sequence we associate an output sequence using the following algorithm. A $0$ at position $l$ is converted to a $1$, or vice versa, only if the digit at position $l+j$ is a $0$ \cite{fn1}. This is a nonlinear problem because the evolution depends on the state of the system, moreover it is a long-range system because the evolution of a digit at site $l$ depends on the digit at a distance $j$. In the classification of elementary cellular automata, and for $j=1$, this corresponds to rule $153$. For $j>1$ we will refer to this rule as long-range rule $153$. 

Steady states of cellular automata can be written exactly using matrix product states \cite{ProsenBuca2017}, however here we are interested in training an MPO that describes the evolution rule of a cellular automata. Such cellular automata rule can be exactly mapped to MPOs \cite{fn4}. 
Hence this is an ideal example to test the functioning of the MPO algorithm. To show how to compute the exact MPO for cellular automata we take the, readily generalizable, example of $j=1$ (see the Appendix \ref{app:automata} for the MPOs for rules $18$ and $30$). In this case, the exact evolution is given by an MPO with bond dimension $D_W=2$. 
For sites $l$ different from $1$ and $L$, the MPOs are 
\begin{align}  
W^{0,0}_{b_{l-1},b_l}=\left[\begin{array}{cc}
0 & 1\\0 & 0
\end{array}\right], \;\;\; W^{0,1}_{b_{l-1},b_l}=\left[\begin{array}{cc}
1 & 0\\0 & 0
\end{array}\right], \nonumber \\   
W^{1,1}_{b_{l-1},b_l}=\left[\begin{array}{cc}
0 & 0\\0 & 1
\end{array}\right], \;\;\; W^{1,0}_{b_{l-1},b_l}=\left[\begin{array}{cc}
0 & 0\\1 & 0
\end{array}\right],     
\end{align}
while for the first and last site they are respectively given by 
\begin{align}  
W^{0,0}_{b_0,b_1}=\left[0,\;1\right],\;\;\; W^{0,1}_{b_0,b_1}=\left[1,\;0\right], \nonumber \\ 
W^{1,1}_{b_0,b_1}=\left[0,\;1\right],\;\;\; W^{1,0}_{b_0,b_1}=\left[1,\;0\right], 
\end{align}
and 
\begin{align}  
W^{0,0}_{b_{L-1},b_L}=\left[\begin{array}{c}
1 \\0
\end{array}\right],\;\;\;\; W^{0,1}_{b_{L-1},b_L}=\left[\begin{array}{c}
0 \\0   
\end{array}\right],\nonumber \\ 
W^{1,1}_{b_{L-1},b_L}=\left[\begin{array}{c}
0 \\1
\end{array}\right],\;\;\;\; W^{1,0}_{b_{L-1},b_L}=\left[\begin{array}{c}
0 \\0   
\end{array}\right]. 
\end{align} 
It is straightforward to check that given an input sequence and an output sequence, the MPO made by the product of the tensors above gives the scalar $1$ only if the output sequence $\vec{\tau}$ is the correct evolution of the input sequence $\vec{\sigma}$, otherwise it gives $0$. It should be noted though, that there are many MPOs which fulfil this requirement, as gauge transformation can be applied to the tensors in a way that their product is unchanged. 

In the following we test our code using an interaction range $j=3$ for which, since the required MPO bond dimension increases as $2^j$, should be described exactly by MPOs with bond dimension $D_W=8$ or above. 

\begin{figure}
\begin{center} 
\includegraphics[width=0.9\columnwidth]{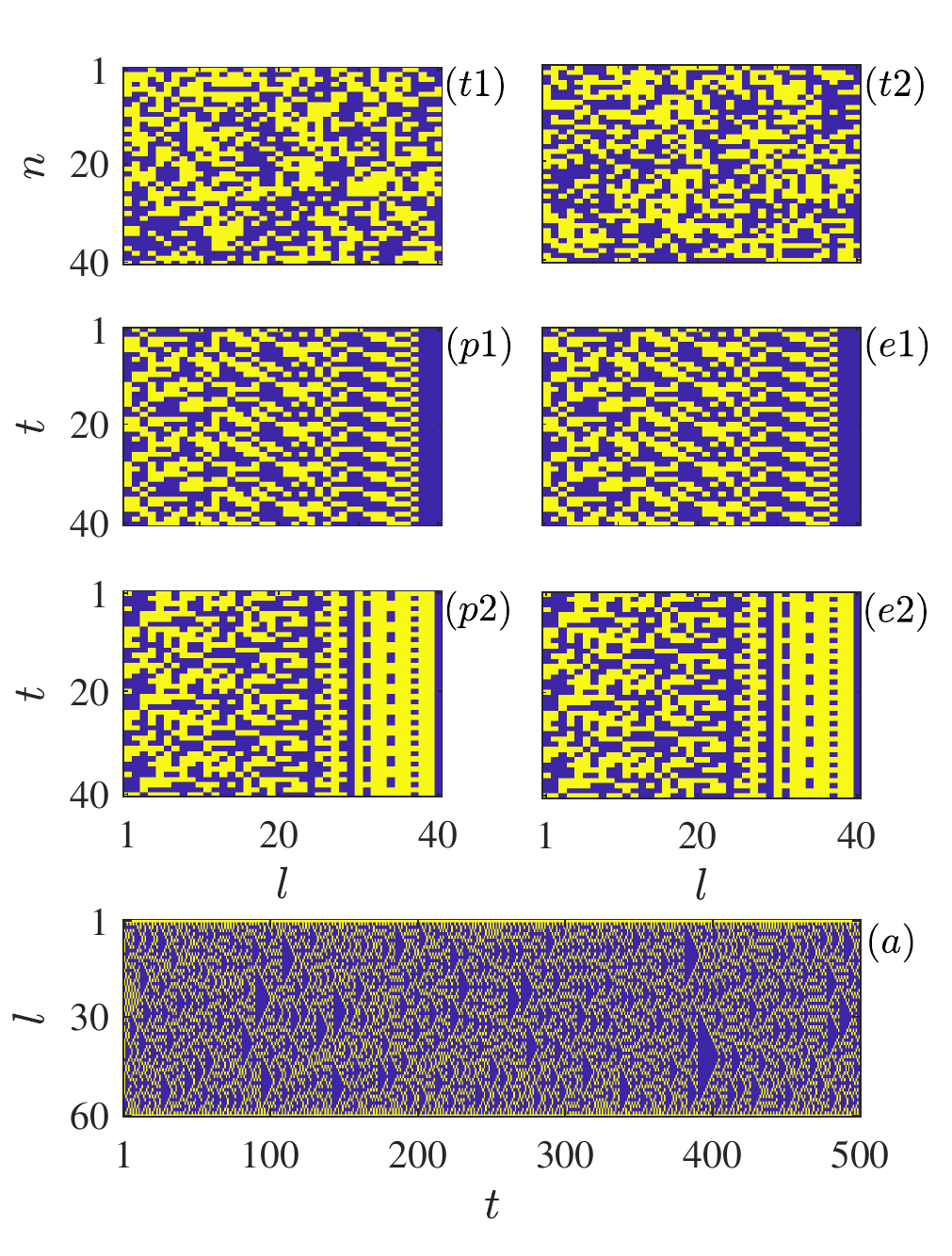}
\caption{Evolution of the long-range rule 153 cellular automata with different random initial conditions from the trained MPO ($p1$, $p2$) and the respective exact evolutions ($e1$, $e2$). Panels (t1-t2) show respectively a sample of $40$ input and output training data. The system is chosen to have $L=40$, interaction distance $j=3$. The training parameters are training sample size $M=5000$, bond dimension of the MPO $D_W=8$, error rate in the training $\epsilon_r=0$, regularizing coefficient $\alpha=0.001$, maximum number of sweeps $\km=20$ and convergence tollerance $\ep_t=10^{-5}$. In panel (a) we show the predicted evolution (which perfectly matches the exact one) for rule 18 with fix boundary conditions. In this case $L=60$, the MPO bond dimension $D_W=4$, and we have used $M=50000$ training pairs of inputs and outputs. Other parameters are $\alpha=0.001$, $\km=20$ and $\epsilon_t=10^{-5}$.  \label{fig:Fig2} }
\end{center}
\end{figure}

To train our MPO we use a set of $M$ input random sequences and their corresponding output. An example of input and output sequences is depicted in Fig.\ref{fig:Fig2}(t1,t2). Within the output sequences we use, with a probability $\epsilon_r$, a wrong sequence also drawn randomly as the input. After having trained our MPO, we consider a set of $N$ initial conditions (which can be different from the ones of the training set) and each one of them we evolve it for $T$ iterations. Two examples of the exact time evolution of sequences are depicted each in one of the two panels (e1) and (e2) of Fig.\ref{fig:Fig2}. The results for the evolution predicted with our MPO algorithm are depicted in Fig.\ref{fig:Fig2}(p1-p2), and show perfect agreement with the exact evolutions. In this case we had considered a system with $L=40$ sites and evolved the initial conditions for $N=40$ steps, while choosing $M=5000$, $D_W=8$, $\al=0.001$, $\km=20$, $\ep_t=10^{-5}$ and $\ep_r=0$ to train the MPO.   
For this case, which is the long-range rule $153$ with $j=3$, the evolution is periodic with a relatively short period. 
It is however interesting to note that, even for evolutions with much longer periods, as for example rule $18$ with fixed boundary conditions for a system with $L=60$ sites, the evolution can be perfectly predicted while no periodic evolution has yet manifested. In fact, in Fig.\ref{fig:Fig2}(a) we show an example of exact prediction of the evolution when using MPO bond dimension $D_W=4$ (which is large enough, as shown in Appendix \ref{app:automata}, to provide an exact evolution of rule $18$), and a number of training input-output pairs $M=50000$.                  
 
\begin{figure}
\begin{center} 
\includegraphics[width=0.8\columnwidth]{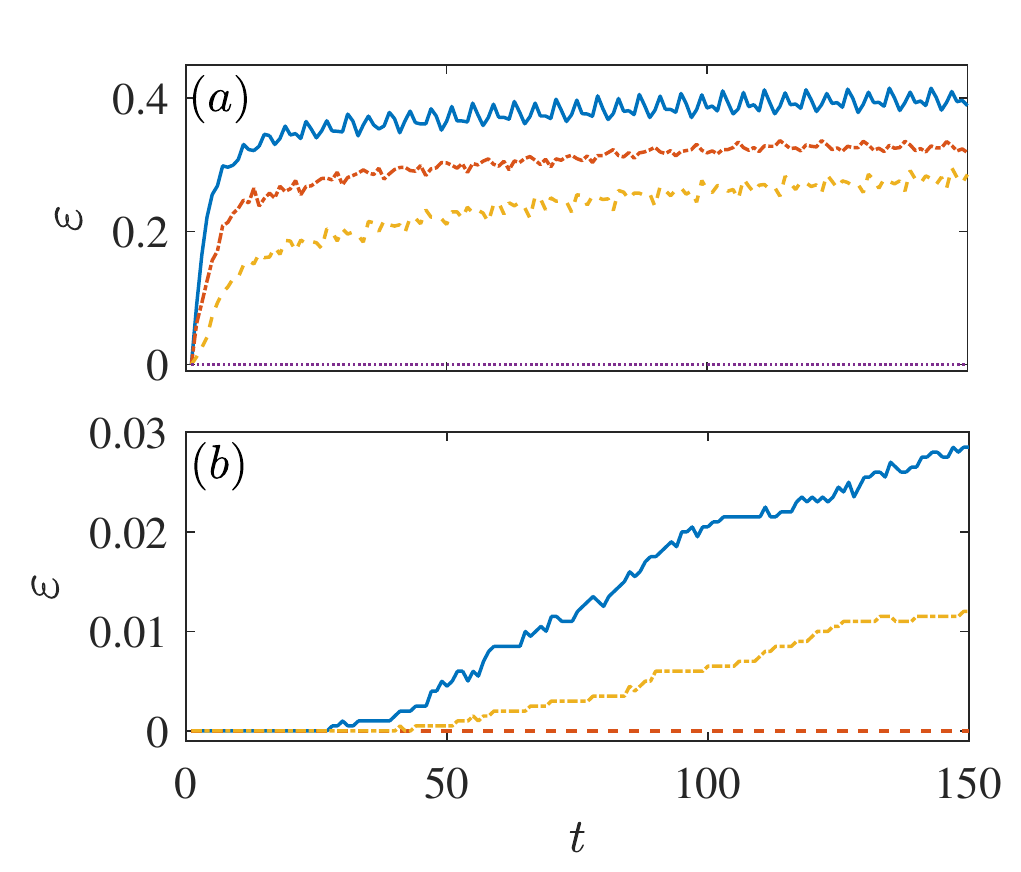}
\caption{Average error $\varepsilon$ between exact and predicted evolution, computed using Eq.(\ref{eq:error}), versus time for different MPO bond dimensions $D_W$ (a), or for different errors $\epsilon_r$ in the training data and sample size $M$ (b). In (a) $M=7000$ while $D_W=5,\;6,\;7$ or $8$ respectively for the blue continuous line, red dot-dashed line, yellow dashed line and purple dotted line. In (b) $D_W=8$ while $M=3000$ and $\epsilon_r=0.2$ for the blue continuous line, $M=7000$ and $\epsilon_r=0.2$ for the red dashed line, and $M=7000$ and $\epsilon_r=0.4$ for the yellow dot-dashed line. Other parameters are $N=100$, $\km=20$, $\epsilon_t=10^{-5}$ and $\alpha=0.001$.  \label{fig:Fig3} }
\end{center}
\end{figure}   
  
We now study the accuracy of the algorithm as a function of the various parameters. We consider again an interaction length $j=3$ and consider $N=100$ initial conditions. In Fig.\ref{fig:Fig3}(a) we show the fraction of wrong predictions versus time averaged over $N$ initial conditions and for different MPO bond dimensions $D_W$. The fraction of wrong predictions is given by 
\begin{align}
\err = \sum_{i,l} |y_{i,l}(t)-\bar{y}_{i,l}(t)|/(L\;N)  \label{eq:error}             
\end{align}       
where $y_{i,l}(t)$ and $\bar{y}_{i,l}(t)$ are respectively the exact and the predicted output for initial condition $i$, position $l$, and time step $t$, where $i\in[1,N]$, $l\in[1,L]$ and $t\in[1,T]$. 
We observe that the error $\err$ decreases as the bond dimension increases, and, as expected, for MPO bond dimension $D_W\ge 8$ the error goes down to $0$ at any time, clearly indicating that the evolution is predicted exactly.  
As we introduce noise in the training samples the accuracy of the prediction decreases, however, for this particular problem, the algorithm still predicts the correct evolution for error rates up to $20\%$ (i.e. $\epsilon_r=0.2$) in the training set provided that the sample size is large enough. This is shown in Fig.\ref{fig:Fig3}(b) where the continuous blue line is computed for $M=3000$ training data, and $\epsilon_r=0.2$, while the red-dashed line is for $M=7000$ and $\epsilon_r=0.2$. In the second case the sample size $M$ is large enough that the prediction is perfect despite the presence of errors. In Fig.\ref{fig:Fig3}(b) we also show that a training data size of $M=7000$, if the error in the training data is even larger, for instance $\epsilon_r=0.4$ as in the yellow dot-dashed line, the error in the evolution $\varepsilon$ increases.         

\begin{figure}
\begin{center}
\includegraphics[width=0.8\columnwidth]{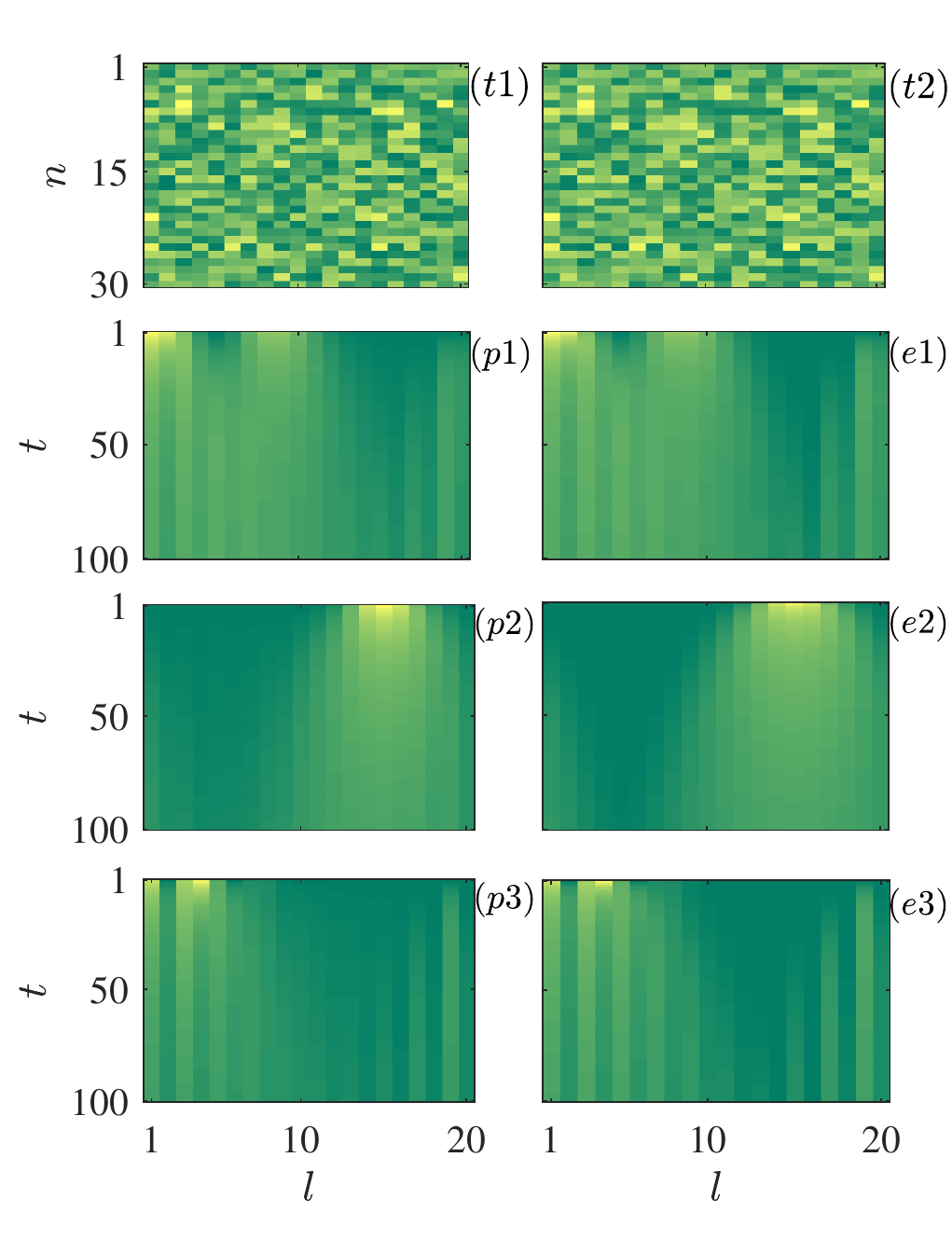}
\caption{Portion of the training data for input (t1) and output (t2). Each of the $n$ rows is an input (t1) or the corresponding output (t2) sequence. The input sequence is chosen from uniformily distributed random numbers between $0$ and $1$ such that their sum is $1$. The output sequence is computed using Eq.(\ref{eq:diffusionequation}). Panels (p1-p3) and (e1-e3) represent the evolution in time of an initial sequence: the left column panels (p1-p3) show the predicted evolution from the MPO algorithm, while the right colum panels (e1-e3) show the exact evolution. The pairs (p1,e1), (p2,e2) and (p3,e3) correspond each to a different initial condition. Other parameters are $D_W=20$, $M=20000$, $\km=20$, $\varepsilon_t=10^{-5}$ and $\alpha=0.001$.     \label{fig:Fig4} }
\end{center}
\end{figure}

\subsubsection{Comparison to conditional random fields}  
The evolution of cellular automata can also be studied with conditional random fields (CRF) \cite{LaffertyPereira2001}. Such method aims to best approximate a conditional probability distribution based on a set of chosen features in the training set (see Appendix \ref{app:crf} for a more detailed description of this method). Such feature can be non-local, thus allowing to take into account long distance correlations (or contextuality), and for this reason the method is used, for example, for natural language processing.  
We find that while the evolution of elementary automata can also be exactly predicted using CRF for the sequence to sequence prediction, there are important differences between the CRF method and the MPO algorithm. For CRF to perform well, one needs to choose the correct features. For example, for the case of rule $153$, one needs to use a bi-gram features based on the value of the local element of the sequence and the element just to its right. For the long-range rule $153$, one should use bi-gram features based on the local element and the one at the particular distance $j$ of the exact evolution. For more general elementary rules, like rule $18$ or rule $30$, one needs to use a tri-gram with features based on the local element and the ones to its right and left. However, in general, it is difficult (or not possible), to have a priori knowledge of the features needed for the algorithm to work well. 

The MPO based algorithm instead has the advantage that, provided the auxiliary dimension $D_W$ is large enough, it can figure which are the most relevant features, whether the element to the right, or to the left, or both, or at a certain, large, distance. This flexibility is what allows MPS and MPO based methods to be so successful in the field of many body quantum system, and in this case, to give perfect prediction with no apriori knowledge of the problem.

\subsection{Coupled nonlinear maps}

After having studied cellular automata, we now consider a coupled nonlinear map. In particular we study a nonlinear diffusive evolution with next-nearest neighbor coupling of a probability distribution, for which the sequences are made of real numbers between $0$ and $1$ whose sum is equal to $1$. In this case, the mapping described in Eq.(\ref{eq:vector}) allows us to convert a certain probability distribution to an MPS of bond dimension $D=1$. In this case, in juxtaposition with the previous study on cellular automata, we use periodic boundary conditions. The evolution is given by 
\begin{align}
P_{l,t+1} &= P_{l,t} \nonumber\\
&+ g_1/2 \left[ \left(P_{l-1,t}\right)^{m_1} +  \left(P_{l+1,t}\right)^{m_1} - 2\left(P_{l,t}\right)^{m_1}\right] \nonumber\\ 
&+ g_2/2 \left[ \left(P_{l-2,t}\right)^{m_2} +  \left(P_{l+2,t}\right)^{m_2} - 2\left(P_{l,t}\right)^{m_2}\right]   \label{eq:diffusionequation}    
\end{align}  
where $t$ is a natural number which represent the time step. The other parameters of the evolution are $g_1$ and $g_2$ which are the diffusion rate to, respectively, the nearest and next nearest sites and $m_1,\;m_2$ are the exponents which make the diffusion nonlinear. 
For the evolution we use initial conditions for which $P_{l,t}\in [0,1]$ up to the times studied.      

Similarly to the case of cellular automata, we train our MPO by choosing random input sequences and their corresponding output sequences. 
In Fig.\ref{fig:Fig4} we show, similarly to Fig.\ref{fig:Fig2}, samples of input and output training data, panels $(t1)$ and $(t2)$, and the comparison between the predicted evolutions, panels $(p1-p3)$, and exact ones, panels $(e1-e3)$, for three different initial conditions (chosen between hundreds which are used for further calculations). 
The initial conditions we use are functions of the position with different number of peaks, each of different variance. This choice is very different from the training data (for which the input sequences were chosen randomly), and also it allows one to have a qualitative understanding of the dynamics. We have in fact chosen the initial conditions as $P_{l,t=1}=( 1+\cos(2 \pi l \lambda/L) )\exp(-((l-l_0)^2)/(2v))/\Gamma $ where $\lambda$ is an integer number uniformly chosen between $1$ and $5$, $l_0$, chosen uniformly between all the site, is the position of a gaussian envelope of standard deviation $\sqrt{v}$ (a real number picked uniformly between $1$ and $5$) and $\Gamma$ is a normalization which ensures that the sum of all values at all the sites gives $1$.  
The evolution of the three different initial conditions in Fig.\ref{fig:Fig2}(p1-p2,e1-e3) shows a peculiar diffusive dynamics.                   

\begin{figure}
\includegraphics[width=\columnwidth]{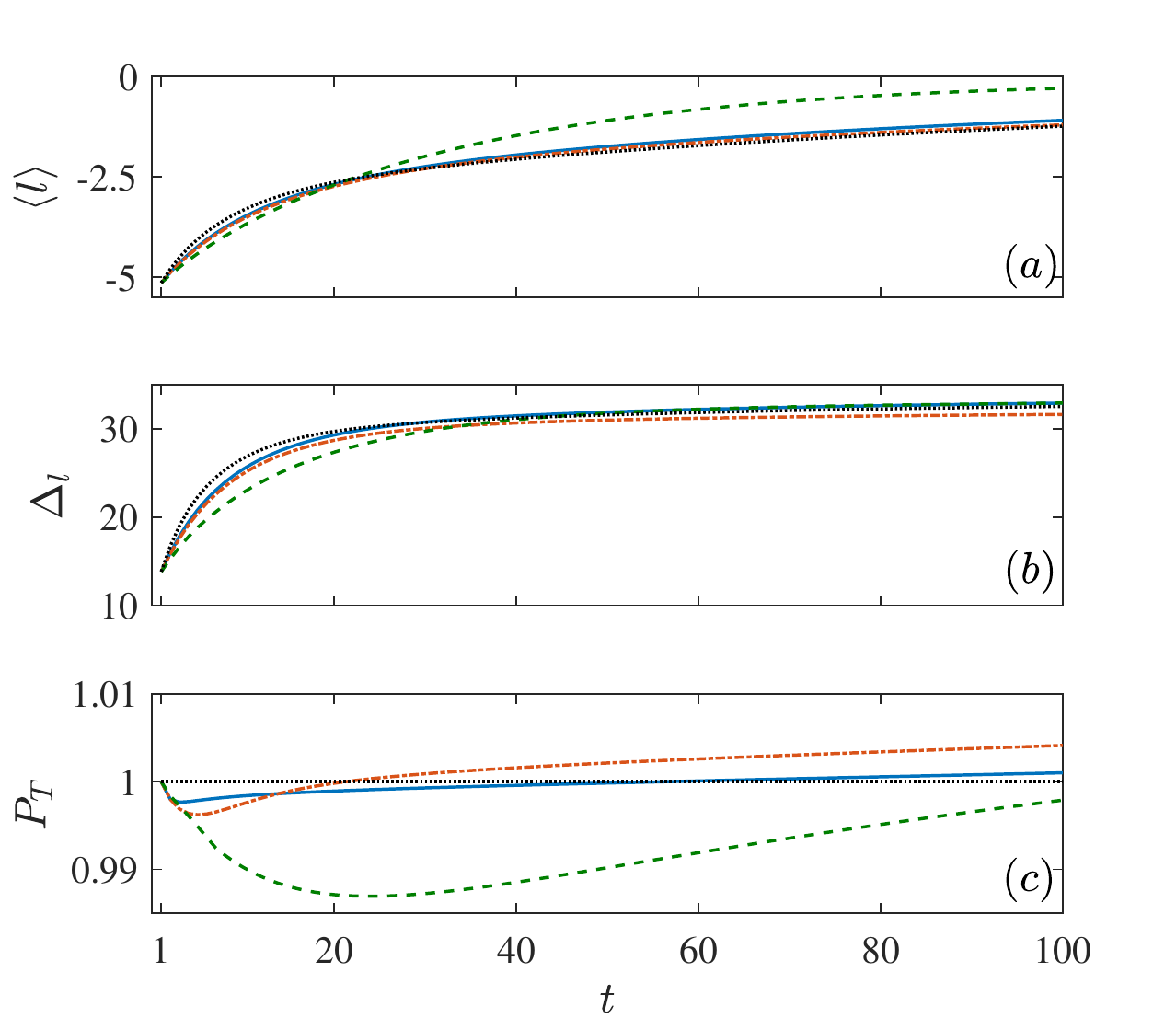}
\caption{(a) Average position $\langle l \rangle$, (b) standard deviation $\Delta_l$ and (c) total probability $P_T$ versus time for $D_W=5$ and $\varepsilon_t=10^{-5}$, green dashed line, $D_W=20$ and $\varepsilon_t=10^{-5}$, red dot-dashed line, and $D_W=20$ and $\varepsilon_t=10^{-7}$, blue continuous line. The exact solutions are represented by the black dotted lines. Other parameters are $N=100$, $M=20000$, $\km=20$ and $\alpha=0.001$.     \label{fig:Fig5} }   
\end{figure}

To gain more insight in the prediction we compute the average position $\langle l \rangle = \sum_l P_{l,t} l$, variance $\Delta_l = \left(\sum_l l^2\; P_{l,t}\right) - \langle l \rangle$ and total probability $P_T = \sum_l P_{l,t}$. These quantities are depicted in Fig.\ref{fig:Fig5}, respectively in panels (a), (b) and (c). In each panel the dotted black lines represent the exact result, the green dashed line the prediction for a bond dimension $D_W=5$ and $\varepsilon_t=10^{-5}$, the red dashed line for $D_W=20$ and $\varepsilon_t=10^{-5}$ while the blue continous line for $D_W=20$ and $\varepsilon_t=10^{-7}$. We observe that the accuracy significantly increases as the bond dimension changes from $D_W=5$ to $D_W=20$ and the norm is better conserved for $\varepsilon_t=10^{-7}$.

While in Fig.\ref{fig:Fig4},\ref{fig:Fig5} we observe that the dynamics is well predicted by the MPO algorithm, to be more quantitative, we compute the average error rate as a function of time for different bond dimensions $D_W$ (panel (a)) and for different number of training data $M$. The results are represented in the two panels of Fig.\ref{fig:Fig6}. In Fig.\ref{fig:Fig6}(a), the bond dimension varies between $D_W=5$ (continuous blue line) to $D_W=10$ (dashed red line) and $D_W=20$ (dot-dashed yellow line), while the size of the training set is kept to $M=20000$. In Fig.\ref{fig:Fig6}(b) instead, we keep the same bond dimension $D_W=20$ and we vary the training data set size between $M=5000$ (continuous blue line), $M=10000$ (dashed red line) and $M=20000$ (dot-dashed yellow line). We observe that the error decreases when both $D_W$ and $M$ increase, and $\varepsilon$ can be as low as $5/1000$.

\subsubsection{Comparison to bidirectional LSTM neural networks}

\begin{figure}
\includegraphics[width=0.9\columnwidth]{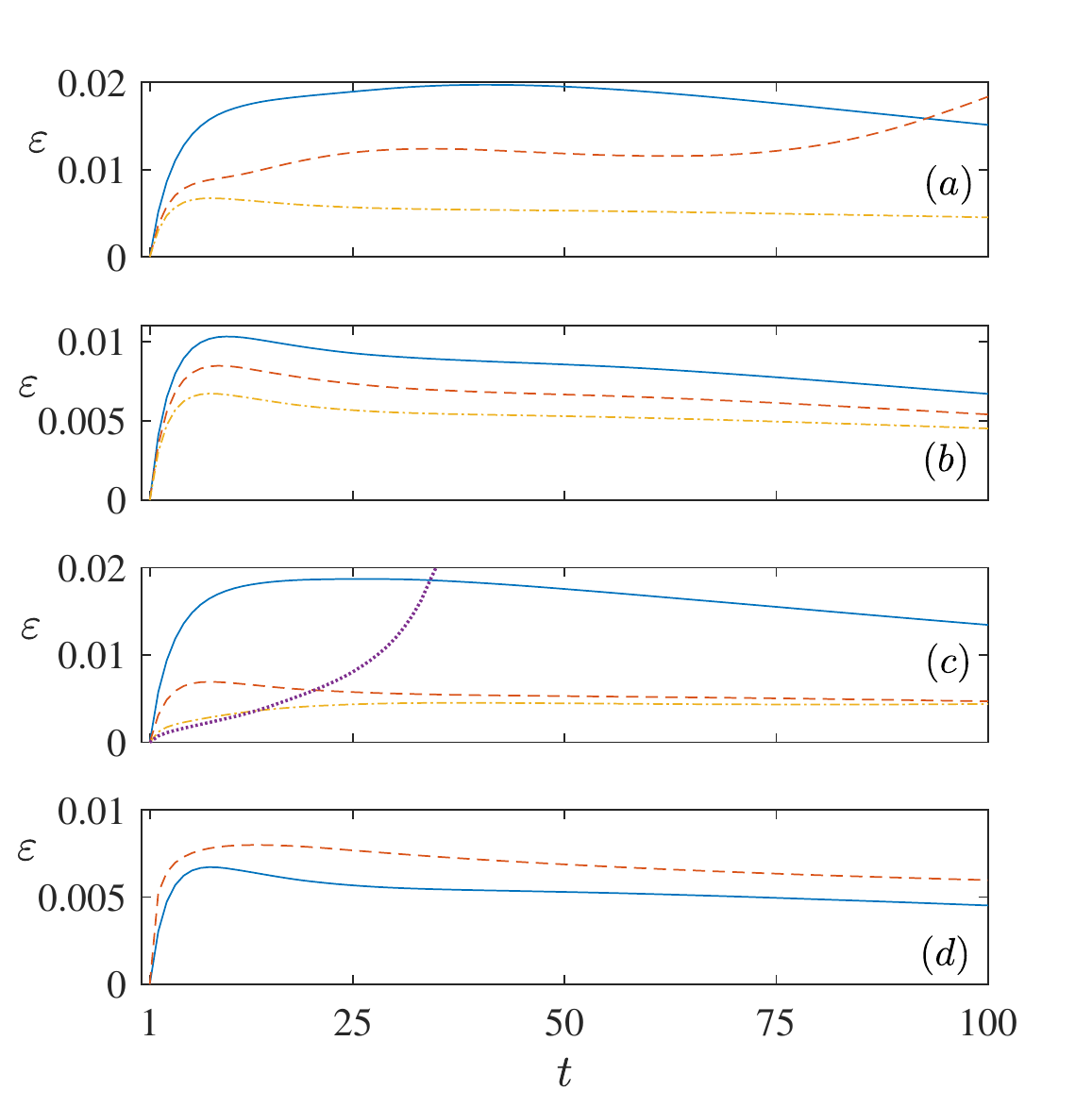}
\caption{Average error $\varepsilon$ between exact and predicted evolution, computed using Eq.(\ref{eq:error}), versus time for different bond dimensions $D$ (a), for different number of training data $M$ (b), or for different regularization constants $\alpha$ (c). In (a) $M=20000$ and $\alpha=0.001$ while $D_W=5,\;10$ or $20$ respectively for the blue continuous line, red dashed line and yellow dot-dashed line. In (b) $D_W=20$ and $\alpha=0.001$ while $M=5000,\;10000$ or $20000$ respectively for the blue continuous line, red dashed line and yellow dot-dashed line. In (c) $D_W=20$ and $M=20000$ while $\alpha=0.1,\;0.001,\;10^{-5}$ or $10^{-7}$ respectively for the blue continuous line, red dashed line, yellow dot-dashed line and purple dotted line. Other parameters are $N=100$, $\km=20$ and $\epsilon_t=10^{-7}$. (d) Comparison with BiLSTM model. Red dashed curve is the error for the BiLSTM model while blue continuous curve is for the MPO algorithm with $D_W=20$, $\alpha=0.001$, $N=100$, $\km=20$ and $\epsilon_t=10^{-7}$. Both algorithms have been trained with $M=20000$ samples and we have used $N=100$ initial conditions. For the BiLSTM model we have used a batch size $d_B=1$, hidden size of LSTM cell $d_{LSTM}=100$ and learning rate $l_R=0.001$.  \label{fig:Fig6} }
\end{figure}

Long short-term memory cells are very useful building blocks for recurrent neural networks \cite{HochreiterSchmidhuber1997, GersCummins2000}. Thanks to their ability of remembering and forgetting information passed to them, they allow to grasp relevant long-range correlations (see Appendix \ref{app:lstm} for a more detailed description of this method). They are thus used for natural language processing, to classify, to predict time series and more. Here we consider a bidirectional network \cite{graves2005framewise}, in which information is passed from left to right and from right to left, with the use of LSTM cells (we will refer to this model as BiLSTM). 
The model is trained by minimizing the distance between the predicted and the exact output. The minimization procedure does not always lower this distance and thus a development set is needed. The role of this set is to check whether the minimization is improving the performance of the model and, in that case, store the current model parameters as the best ones. 
More precisely, the training is done in this way: We use one input-output pair at the time to train the model with Adam algorithm \cite{kingba2014}. After doing this $10000$ times, we test the current model parameters with the development set, and record the model parameters only if the performance on the development set is better than the previous ones. We iterate this procedure for $100$ epochs (one epoch corresponds to using once the complete training data set). 
In our realization we have used $20000$ training input-output pairs and another $20000$ for the development set. We have used different combinations of the hyper-parameters hidden size of LSTM cells $d_{LSTM}$ (we considered $d_{LSTM}=50$, $100$ and $200$), and batch size $d_B$ (we considered $d_B=1$, $10$, $25$ and $50$) and picked those which performed best on the development set ($d_{LSTM}=100$ and $d_B=1$). More details can be found in the SI. For the optimization Adam algorithm we use a learning rate $l_R=0.001$. 
The performance of the BiLSTM model (red-dashed line) and the MPO algorithm (blue continuous line) are compared in Fig.\ref{fig:Fig6}(d), where the error $\varepsilon$ is plotted against time. We note that in these realizations, in which the total number of parameters is comparable in the two models, the performance of the MPO algorithm improve on those of the BiLSTM.  

\subsection{Classification}         
It has been previously shown that algorithms based on MPS can be useful for classification tasks \cite{NovikovOseledets2017, StoudenmireSchwab2016, Stoudenmire2017}. It is straightforward to show that the MPO algorithm can also be used for such purpose. In order to explain how to do so effectively, we consider a well known example which is that of categorizing pictures of handwritten digits from the MNIST data set. In this problem, handwritten digits from $0$ to $9$ are converted in gray scale pictures with $28\times 28$ pixels, and the data set is composed of $60000$ training images and $10000$ validation images (development set), each associated to one of the ten possible digits. In order to threat this problem we convert each gray scale figure to a sequence of real numbers between $0$ and $1$ by evaluating how dark each pixel is. Note that this portion of the treatment of the problem is completely analogous to \cite{StoudenmireSchwab2016}. From these input sequences we generate input MPSs which will be used for the training of the MPO. The output is instead composed of only ten possible digits, while our algorithm can return one of the $d^{L}$ possible sequences, where $d=2$. We then write each possible output as a one hot vector of size $L$ with the care that the element of the vector which is equal to $1$ is in positions of the vector which span, more or less evenly, between the first and the last site \cite{fn2, fn3}. By doing so, we have ten possible different vector output, which we can readily write as bond dimension $D=1$ MPS, as in the previous example for cellular automata. We shall stress that in this problem the input sequence is composed of a real numbers while the output is made of integer numbers. This exemplifies that the different nature of the input and output sequences is not an obstacle to the functioning of the algorithm. 

We then use the MNIST training data set ($60000$ images) and the development set ($10000$ images) \cite{ChristopherCortes}. We find that by using a bond dimension $D=10$ we obtain $93.9\%$ accuracy on the training data set and $94.0\%$ on the development set. For $D=20$ the results improve to $97.6\%$ accuracy for the training data set and $97.2\%$ accuracy in the testing. In both cases we have used regularizing coefficient $\alpha=0.001$ and maximum number of sweeps $\km=10$. These results are comparable to \cite{StoudenmireSchwab2016}.

\section{Conclusions}\label{sec:conclusions}       
We have presented an algorithm based on matrix product operators, MPOs, for sequence to sequence prediction. In this work we have presented the main algorithm and, to show its versatility and effectiveness, we have applied it to various examples ranging from evolution of cellular automata, nonlinear coupled maps and also for classification. 

Two important aspects of using MPOs, and hence MPS for learning should be stressed now. First, the code can accept input and output sequences which are probabilistic. In this case, the input and output probabilities are each converted to matrix product states with a bond dimension typically larger than $1$. These input and output matrix product states can then be readily used to train the MPO. The other noteworthy aspect, which should be further investigated, is that the algorithm does not return a mere output sequence, but a matrix product state which is then converted to an output sequence. Such matrix product state contains information on the building up of correlations during the evolution and it can thus be used to learn more about the system.  

Our comparison of the MPO algorithm with CRF shows that the MPO algorithm does not require a-priori knowledge about the relevant features which are important to describe the conditional probabilities, but that it is able to find them or approximate them at best, within the limitation of the size of the matrices used. For the tasks studied the predictions of the MPO algorithm where comparable in accuracy to bidirection LSTM model, however the MPO algorithm was faster. Furthermore, the speed of the current version of the MPO algorithm can be significantly boosted.       

In the future we aim to extend and apply such algorithm to problems related to natural language processing, as these problems are inherently one dimensional, a situation for which matrix product states and operators algorithms perform at their best. 

Various improvements are required such as allowing for the study of sequences of unknown lengths or for which the length of the input is different from that of the output. Fundamental questions regarding the effectiveness of training an MPO depending on the underlying dynamics also need to be addressed. These will be the subjects of follow-up works.

\section*{Acknowledgments} 
We acknowledge support from the SUTD-MIT International Design Centre. D.P. acknowledges fruitful discussions with F.C. Binder, S. Lin and N. Lim.

\newpage 
\clearpage
\newpage 
\begin{appendix}  
\renewcommand{\thefigure}{A\arabic{figure}}
\setcounter{figure}{0}

\section{Iterative minimization} \label{app:minimization}
\begin{figure}
\includegraphics[width=0.9\columnwidth]{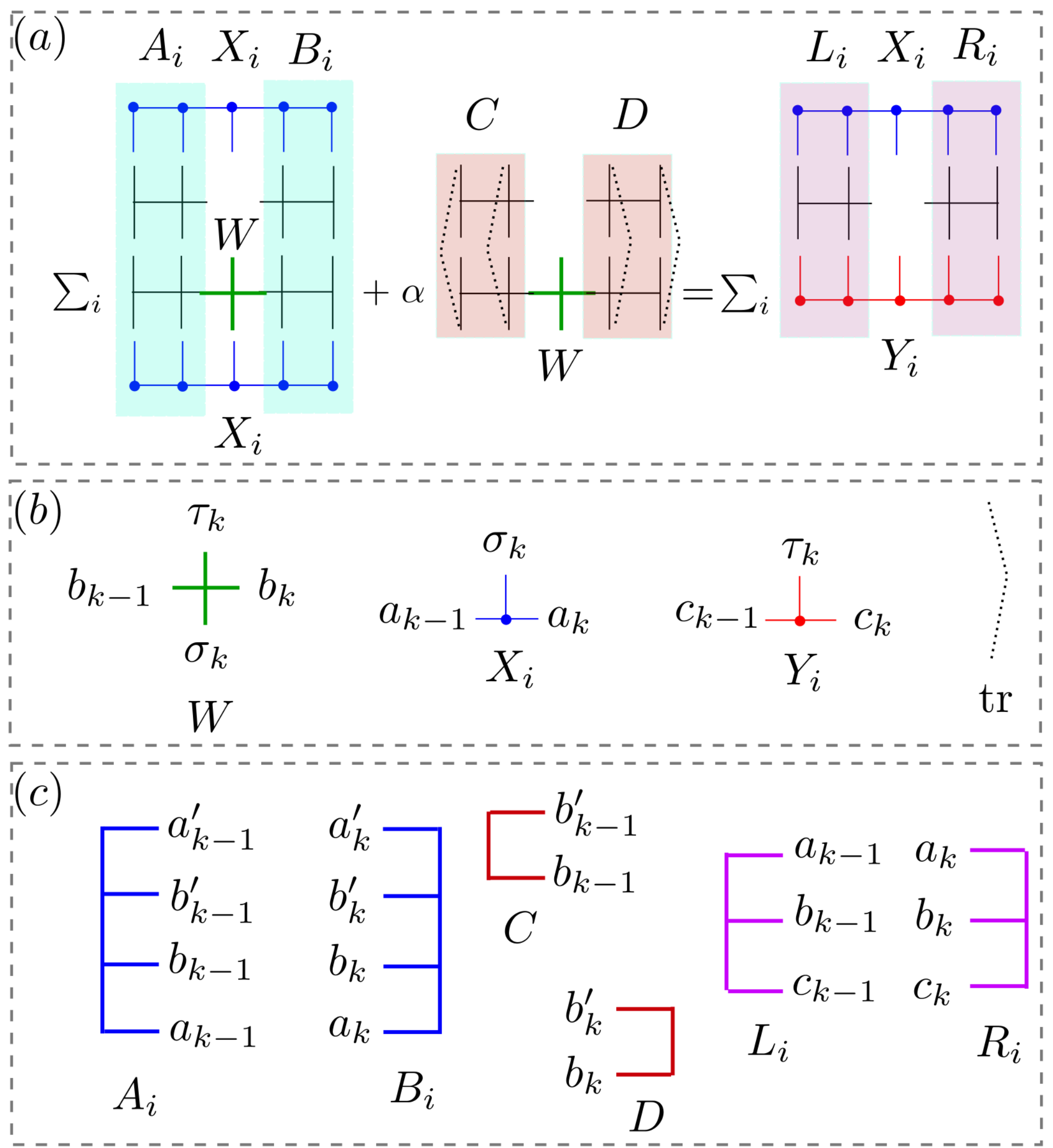}
\caption{(a) Graphical representation of the linear equation (\ref{eq:solve}) in which every tensor is a structure with a different number of `legs' depending on the number of tensor indices (for lighter notation we have removed the indices $l$, $\sigma_l$ and $\tau_l$). The vertical lines correspond to physical indices (like $\sigma_l$ or $\tau_l$) while the horizontal lines to auxiliary indices (like $a_l$, $b_l$ or $c_l$). Joined legs are summed over. In panel (b) we show the elementary tensors $W_{b_{l-1}, b_l}^{\sigma_l, \ta_l}$, $X_{i, a_{l-1}, a_l}^{\sigma_l}$ and $Y_{i, c_{l-1}, c_l}^{\ta_l}$ while in panel (c) we show the graphical representation of intermediate tensors which are computed to evaluate more efficiently Eq.(\ref{eq:solve}). \label{fig:Fig1A}}
\end{figure}

Here we describe in more detail how we use Eq.(\ref{eq:minimize}) to compute the optimal $W^{\s_l,\ta_l}_{b_{l-1},b_l}$. As commonly done in many MPS algorithms for quantum mechanical systems, it is convenient to group the product of the tensors in various parts as depicted in Fig.\ref{fig:Fig1A}. 
From the simple derivative in Eq.(\ref{eq:minimize}), the tensor $W^{\s_l,\ta_l}_{b_{l-1},b_l}$ will be the result of the linear equation 
\begin{align}
\sum_{b_{l-1}, b_l, \sigma_l} & \left( F_{b_{l-1}, b_{l}, b_{l-1}^{\prime}, b_l^{\prime}}^{\sigma_l, \sigma_l^{\prime}} + \alpha C_{b_{l-1}, b_{l-1}^{\prime}}D_{b_{l}, b_{l}^{\prime}} \right)W_{b_{l-1}, b_l}^{\sigma_l, \ta_l} \nonumber\\ &= U_{b_{l-1}^{\prime}, b^{\prime}_{l}}^{\sigma_l^{\prime}, \ta_l}. \label{eq:solve}
\end{align}    
This equation is described graphically in Fig.\ref{fig:Fig1A}(a) where the elementary tensors $W_{b_{l-1}, b_l}^{\sigma_l, \ta_l}$, $X_{i, a_{l-1}, a_l}^{\sigma_l}$ and $Y_{i, c_{l-1}, c_l}^{\ta_l}$ are described in Fig.\ref{fig:Fig1A}(b) and where the dotted line is used to indicate the trace operation. The tensor $F^{\s_l,\s'_l}_{b_{l-1},b_l,b'_{l-1},b'_l}$ is composed of the product of tensors to the left and to the right of the site of interest    
\begin{align}
F_{b_{l-1}, b_{l}, b_{l-1}^{\prime}, b_l^{\prime}}^{\sigma_l, \sigma_l^{\prime}} =& 
\sum_{\substack{i, a_{l-1},\\ c_{l-1}, a_l, c_l}} A_{i, a_{l-1}, b_{l-1}, b_{l-1}^{\prime}, c_{l-1}} \nonumber \\  
&\times B_{i, a_l, b_l, b_l^{\prime}, c_l} X_{i, a_{l-1}, a_l}^{\sigma_l}X_{i, c_{l-1}, c_l}^{\sigma_l^{\prime}} \label{eq:Ftensor}   
\end{align}  
with 
\begin{align}
&A_{i, a_{l-1}, b_{l-1}, b_{l-1}^{\prime}, a_{l-1}^{\prime}} =  \nonumber \\  & \sum_{\substack{a_k, b_k, b_k^{\prime}, a_k^{\prime},\\ \sigma_k, \ta_k, \sigma_k^{\prime}}}  \prod_{k <l} \left( W_{b_{k-1}, b_k}^{\sigma_k, \ta_k}W_{b_{k-1}^{\prime}, b_k^{\prime}}^{\sigma_k^{\prime}, \ta_k} X_{i, a_{k-1}, a_k}^{\sigma_k} X_{i, a_{k-1}^{\prime}, a_k^{\prime} }^{\sigma_k^{\prime}} \right) \label{eq:Atensor}  
\end{align}
and 
\begin{align}
&B_{i, a_{l}, b_{l}, b_{l}^{\prime}, a_{l}^{\prime}} =  \nonumber \\  & \sum_{\substack{a_k, b_k, b_k^{\prime}, a_k^{\prime},\\ \sigma_k, \ta_k, \sigma_k^{\prime}}}  \prod_{k >l} \left( W_{b_{k-1}, b_k}^{\sigma_k, \ta_k}W_{b_{k-1}^{\prime}, b_k^{\prime}}^{\sigma_k^{\prime}, \ta_k} X_{i, a_{k-1}, a_k}^{\sigma_k} X_{i, a_{k-1}^{\prime}, a_k^{\prime} }^{\sigma_k^{\prime}} \right) \label{eq:Btensor}  
\end{align}
and where $A_{i,a_0,b_0,b'_0,a'_0}=A_{i,a_L,b_L,b'_L,a'_L}=1$. These tensors are depicted in Fig.\ref{fig:Fig1A}(c). 
Note also that here and in the following, the sum over the tensor indices are done only if the same index label appears in two different tensors. From the graphical representation of our equations in Fig.\ref{fig:Fig1A}, this would mean that the sum is done over the indices for lines joining different tensors.    
For Eq.(\ref{eq:solve}) we also need to compute the tensors $C$, $D$ and $U$. Tensor $U$ is given by 
\begin{align}
& U_{b_{l-1}^{\prime}, b_{l}^{\prime}}^{\sigma_l^{\prime}, \ta_l} = \nonumber \\ 
& \sum_{ \substack{ i,a_{l-1}^{\prime}, c_{l-1},\\ a_l^{\prime}, c_l}} L_{i, a_{l-1}^{\prime}, b_{l-1}^{\prime}, c_{l-1}}  R_{i, a_l^{\prime}, b_l^{\prime}, c_l} X_{i, a_{l-1}^{\prime}, a_l^{\prime}}^{\sigma_l^{\prime}} Y_{i, c_{l-1}, c_l}^{\ta_l} \label{eq:Utensor}
\end{align}   
where 
\begin{align}
& L_{i, a_{l-1}^{\prime}, b_{l-1}^{\prime}, c_{l-1}} =  
 \sum_{ \substack{ a_k^{\prime}, b_k^{\prime}, c_k \\ \sigma_k, \ta_k}} \prod_{k <l} \left( W_{b_{k-1}^{\prime}, b_k^{\prime}}^{\sigma_k, \ta_k}X_{i, a_{k-1}^{\prime}, a_k^{\prime}}^{\sigma_k} Y_{i, c_{k-1}, c_k}^{\ta_k} \right)  \label{eq:Ltensor}     
\end{align} 
and 
\begin{align}
& R_{i, a_l^{\prime}, b_l^{\prime}, c_l} =  
 \sum_{ \substack{ a_k^{\prime}, b_k^{\prime}, c_k \\ \sigma_k, \ta_k}} \prod_{k > l} \left( W_{b_{k-1}^{\prime}, b_k^{\prime}}^{\sigma_k, \ta_k}X_{i, a_{k-1}^{\prime}, a_k^{\prime}}^{\sigma_k} Y_{i, c_{k-1}, c_k}^{\ta_k} \right).  \label{eq:Rtensor}     
\end{align} 
The $U$ tensor is depicted in the right-hand side of the equation in Fig.\ref{fig:Fig1A}(a). 
$C$ and $D$, are instead given by      
\begin{align}
C_{b_{l-1}, b'_{l-1}} =  
 \sum_{ \substack{ b_k, b'_k,  \\ \sigma_k, \ta_k }} \prod_{k < l} \left( W_{b_{k-1}, b_k}^{\sigma_k, \ta_k} W_{b'_{k-1}, b'_k}^{\sigma_k, \ta_k} \right)  \label{eq:Ctensor} 
\end{align}
and 
\begin{align}
D_{b_{l}, b'_{l}} =  
 \sum_{ \substack{ b_k, b'_k,  \\ \sigma_k, \ta_k }} \prod_{k > l} \left( W_{b_{k-1}, b_k}^{\sigma_k, \ta_k} W_{b'_{k-1}, b'_k}^{\sigma_k, \ta_k} \right).  \label{eq:Dtensor} 
\end{align}      
Computing the products of the tensors in such groupings allows to speed up the calculations of the multiplications of tensors by growing iteratively  these block of tensors. Note also that the number of tensors required for these calculations scales linearly with the number of training data $N$. The memory requirements can be mitigated by computing the products when needed.

\section{Example of matrix product operator rewriting of cellular automata evolution} \label{app:automata}
The cellular automata considered in the main text, rule $153$, is such that the value at one site at the next iteration only depends on a site to its right and it is independent from those to the left. For this reason the evolution can be exactly described by a matrix product operator of bond dimension $D=2$. However, in general, the $256$ rules of cellular automata by Wolfram \cite{Wolfram1983}, can be exactly described by an MPO of bond dimension $D=4$. We here describe two examples, rules $18$ and $30$. Rule $18$ is also used in the main text as, also for fixed boundary conditions, can produce evolutions with long periods. Rule $30$ instead is chaotic, but not within fix boundary conditions considered here, and hence, with such boundary conditions the evolution becomes regular in a short time (periodic boundary conditions can be also considered with MPOs but in this case the bond dimension $D_W$ could be the square of the fix boundaries condition case). 

Since for both cases, rule $18$ and rule $30$, the maximum needed bond dimension is $D=4$, there will be $5$ types of tensors. Those on the first or last site, those on the second and second last site and the other tensors in the middle of the chain. 
For rule $30$ we have, for the first site   
\begin{align}  
W^{0,0}_{b_0,b_1}=\left[1\;0\right], \;\;\; W^{0,1}_{b_{0},b_1}=\left[0\;0\right], \nonumber \\   
W^{1,0}_{b_{0},b_1}=\left[0\;0\right], \;\;\; W^{1,1}_{b_{0},b_1}=\left[0\;1\right],     
\end{align}
for the second site, 
\begin{align}  
W^{0,0}_{b_{1},b_2}=\left[\begin{array}{cccc}
1 & 0 & 0 & 0\\ 0 & 1 & 0 & 0
\end{array}\right], \;\;\; W^{0,1}_{b_{1},b_2}=\left[\begin{array}{cccc}
0 & 1 & 0 & 0\\ 1 & 0 & 0 & 0
\end{array}\right], \nonumber \\   
W^{1,0}_{b_{1},b_2}=\left[\begin{array}{cccc}
0 & 0 & 0 & 0\\ 0 & 0 & 1 & 1
\end{array}\right], \;\;\; W^{1,1}_{b_{1},b_2}=\left[\begin{array}{cccc}
0 & 0 & 1 & 1\\ 0 & 0 & 0 & 0
\end{array}\right],     
\end{align}
for the intermediate sites 
\begin{align}  
W^{0,0}_{b_{l-1},b_l}=\left[\begin{array}{cccc}
1 & 0 & 0 & 0\\ 0 & 0 & 0 & 0 \\ 0 & 1 & 0 & 0\\ 0 & 0 & 0 & 0 
\end{array}\right], \;\;\; W^{0,1}_{b_{l-1},b_l}=\left[\begin{array}{cccc}
0 & 1 & 0 & 0\\ 0 & 0 & 0 & 0 \\ 1 & 0 & 0 & 0\\ 0 & 0 & 0 & 0 
\end{array}\right], \nonumber \\   
W^{1,1}_{b_{l-1},b_l}=\left[\begin{array}{cccc}
0 & 0 & 0 & 0\\ 0 & 0 & 0 & 0 \\ 0 & 0 & 0 & 0\\ 0 & 0 & 1 & 1 
\end{array}\right], \;\;\; W^{1,0}_{b_{l-1},b_l}=\left[\begin{array}{cccc}
0 & 0 & 0 & 0\\ 0 & 0 & 1 & 1 \\ 0 & 0 & 0 & 0\\ 0 & 0 & 0 & 0 
\end{array}\right],     
\end{align}
for the before last site 
\begin{align}  
W^{0,0}_{b_{L-2},b_{L-1}}=\left[\begin{array}{cc}
1 & 0 \\ 0 & 0 \\ 0 & 1\\ 0 & 0 
\end{array}\right], \;\;\; W^{0,1}_{b_{L-2},b_{L-1}}=\left[\begin{array}{cc}
0 & 1\\ 0 & 0 \\ 1 & 0\\ 0 & 0 
\end{array}\right], \nonumber \\   
W^{1,1}_{b_{L-2},b_{L-1}}=\left[\begin{array}{cc}
0 & 0 \\ 0 & 0 \\ 0 & 0 \\ 1 & 1 
\end{array}\right], \;\;\; W^{1,0}_{b_{L-2},b_{L-1}}=\left[\begin{array}{cc}
0 & 0 \\ 1 & 1 \\ 0 & 0\\ 0 & 0 
\end{array}\right],   
\end{align}
and for the last site 
\begin{align}  
W^{0,0}_{b_{L-1},b_L}=\left[\begin{array}{c}
1\\0 
\end{array}\right], \;\;\; W^{0,1}_{b_{L-1},b_L}=\left[\begin{array}{c}
0\\ 0
\end{array}\right], \nonumber \\   
W^{1,0}_{b_{L-1},b_L}=\left[\begin{array}{c}
0\\0 
\end{array}\right], \;\;\; W^{1,1}_{b_{L-1},b_L}=\left[\begin{array}{c}
0 \\1 
\end{array}\right].      
\end{align}

For rule $18$ we have, for the first site   
\begin{align}  
W^{0,0}_{b_0,b_1}=\left[1\;0\right], \;\;\; W^{0,1}_{b_{0},b_1}=\left[0\;0\right], \nonumber \\   
W^{1,0}_{b_{0},b_1}=\left[0\;0\right], \;\;\; W^{1,1}_{b_{0},b_1}=\left[1\;1\right],     
\end{align}
for the second site, 
\begin{align}  
W^{0,0}_{b_{1},b_2}=\left[\begin{array}{cccc}
1 & 0 & 0 & 0\\ 0 & 1 & 0 & 0
\end{array}\right], \;\;\; W^{0,1}_{b_{1},b_2}=\left[\begin{array}{cccc}
0 & 1 & 0 & 0\\ 1 & 0 & 0 & 0
\end{array}\right], \nonumber \\   
W^{1,0}_{b_{1},b_2}=\left[\begin{array}{cccc}
0 & 0 & 1 & 1\\ 0 & 0 & 1 & 1
\end{array}\right], \;\;\; W^{1,1}_{b_{1},b_2}=\left[\begin{array}{cccc}
0 & 0 & 0 & 0\\ 0 & 0 & 0 & 0
\end{array}\right],     
\end{align}
for the intermediate sites 
\begin{align}  
W^{0,0}_{b_{l-1},b_l}=\left[\begin{array}{cccc}
1 & 0 & 0 & 0\\ 0 & 0 & 0 & 0 \\ 0 & 1 & 0 & 0\\ 0 & 0 & 0 & 0 
\end{array}\right], \;\;\; W^{0,1}_{b_{l-1},b_l}=\left[\begin{array}{cccc}
0 & 1 & 0 & 0\\ 0 & 0 & 0 & 0 \\ 1 & 0 & 0 & 0\\ 0 & 0 & 0 & 0 
\end{array}\right], \nonumber \\   
W^{1,1}_{b_{l-1},b_l}=\left[\begin{array}{cccc}
0 & 0 & 0 & 0\\ 0 & 0 & 1 & 1 \\ 0 & 0 & 0 & 0\\ 0 & 0 & 1 & 1 
\end{array}\right], \;\;\; W^{1,0}_{b_{l-1},b_l}=\left[\begin{array}{cccc}
0 & 0 & 0 & 0\\ 0 & 0 & 0 & 0 \\ 0 & 0 & 0 & 0\\ 0 & 0 & 0 & 0 
\end{array}\right],     
\end{align}
for the before last site 
\begin{align}  
W^{0,0}_{b_{L-2},b_{L-1}}=\left[\begin{array}{cc}
1 & 0 \\ 0 & 0 \\ 0 & 1\\ 0 & 0 
\end{array}\right], \;\;\; W^{0,1}_{b_{L-2},b_{L-1}}=\left[\begin{array}{cc}
0 & 1\\ 0 & 0 \\ 1 & 0\\ 0 & 0 
\end{array}\right], \nonumber \\   
W^{1,1}_{b_{L-2},b_{L-1}}=\left[\begin{array}{cc}
0 & 0 \\ 1 & 1 \\ 0 & 0 \\ 1 & 1 
\end{array}\right], \;\;\; W^{1,0}_{b_{L-2},b_{L-1}}=\left[\begin{array}{cc}
0 & 0 \\ 0 & 0 \\ 0 & 0\\ 0 & 0 
\end{array}\right],     
\end{align}
and for the last site 
\begin{align}  
W^{0,0}_{b_{L-1},b_L}=\left[\begin{array}{c}
1\\0 
\end{array}\right], \;\;\; W^{0,1}_{b_{L-1},b_L}=\left[\begin{array}{c}
0\\ 0
\end{array}\right], \nonumber \\   
W^{1,0}_{b_{L-1},b_L}=\left[\begin{array}{c}
0\\0 
\end{array}\right], \;\;\; W^{1,1}_{b_{L-1},b_L}=\left[\begin{array}{c}
0 \\1 
\end{array}\right].      
\end{align}

\section{Conditional random fields} \label{app:crf}

\textit{Conditional random fields} (CRF)~\cite{LaffertyPereira2001} is a popular model for structured prediction and has been widely used in natural language processing tasks such as part-of-speech (POS) tagging and named entity recognition~\cite{ma2016end}.

Given an input sequence $\vec{x} = \left(x_1, x_2, \cdots, x_L\right)$, where $L$ is the number of elements, the probability of predicting a possible output sequence $\vec{y} = \left(y_1, y_2, \cdots, y_L\right)$ is defined as       
\begin{align}
p(\vec{y}_i | \vec{x}_i) 
& = 
\frac{\exp(   \vec{w}^{\mathbf{T}}      \vec{f}(\vec{x}_i, \vec{y}_i ) )     }
{ Z(\vec{x}) }\\
Z(\vec{x}_i) & =
\sum_{\vec{y}_k }  \exp(      \vec{w}^{\mathbf{T}}      \vec{f}(\vec{x}_i, \vec{y}_k ) )
\end{align}
where  $ \vec{f}(\vec{x}_i, \vec{y}_i  )$ is the feature vector (carefully chosen by the user depending on the problem), $\vec{w}$ is the weight vector consisting of parameters for the features, and $Z(\vec{x}_i)$ is the partition function used for normalization. $\mathbf{T}$ stands for transpose of a matrix/vector. 
The number of parameters is the number of features, which is the size of the feature vector. 
We aim to minimize the negative log-likelihood with $L_2$ regularization 
\begin{equation}
\mathcal{L}(\vec{w}) = -\sum_{i}\log p(\vec{y}_{i} | \vec{x}_{i}) + \lambda \vec{w}^{\mathbf{T}} \vec{w}
\end{equation}
where $(\vec{x}_{i}, \vec{y}_{i})$ is the $i$-th training instance and $\lambda$ is the $L_2$ regularization coefficient. 
Since the objective function is convex, we can make use of the L-BFGS~\cite{liu1989limited} algorithm to optimize it. 
The gradient with respect to each parameter $w_k$ is calculated by setting
\begin{align}
\frac{\partial \mathcal{L}}{\partial w_k}=0
\end{align}

\section{Bidirectional long short-term memory network(LSTM)} \label{app:lstm}

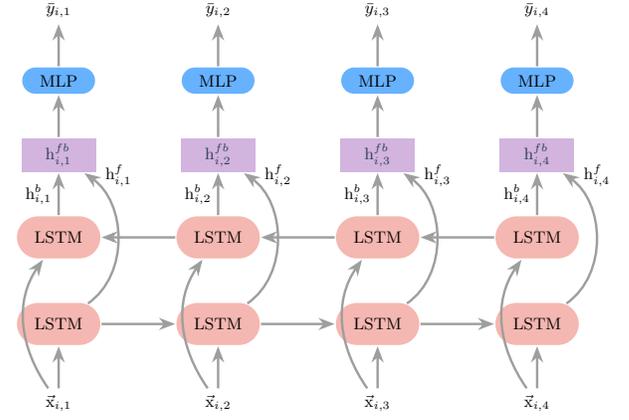
\begin{figure}[t!]
	\centering
	\scalebox{0.7}{
	\begin{tikzpicture}[node distance=8.0mm and 20.0mm, >=Stealth,
	lstm/.style={draw=none, minimum height=8mm,rounded rectangle, fill=lowred, minimum width=18mm, text=fontgrey, label={center:LSTM}},
	textNode/.style={draw=none, rectangle},
	hidden/.style={draw=none, minimum height=3mm, rectangle, fill=lowpurple, minimum width=14mm, text=fontgrey},
	mlp/.style={draw=none, minimum height=5mm, rounded rectangle, fill=lowblue, minimum width=16mm, text=fontgrey, label={center:MLP}},
	emb/.style={draw=black, rectangle, fill=embcolor,blur shadow={shadow blur steps=5,shadow blur extra rounding=1.3pt}},
	chainLine/.style={line width=1.3pt,->, color=mygrey},
	curve/.style={line width=1.3pt, color=mygrey},
	line width=1.3pt
	]

	\node[lstm, yshift=5mm] at (-1.5cm, -5cm) (g1) {};
	\node[lstm, right=of g1, xshift=-6mm](g2) {};
	\node[lstm, right=of g2, xshift=-6mm](g3) {};
	\node[lstm, right=of g3, xshift=-6mm](g4) {};
	\draw [chainLine] (g1) to (g2);
	\draw [chainLine] (g2) to (g3);
	\draw [chainLine] (g3) to (g4);
	\node[textNode, below=of g1] (x1) {$\vec{\text{x}}_{i,1}$};
	\node[textNode, below=of g2] (x2) {$\vec{\text{x}}_{i,2}$};
	\node[textNode, below=of g3] (x3) {$\vec{\text{x}}_{i,3}$};
	\node[textNode, below=of g4] (x4) {$\vec{\text{x}}_{i,4}$};
	\draw [chainLine] (x1) to (g1);
	\draw [chainLine] (x2) to (g2);
	\draw [chainLine] (x3) to (g3);
	\draw [chainLine] (x4) to (g4);
	
	\node[lstm, above=of g1] (b1) {};
	\node[lstm, above=of g2] (b2) {};
	\node[lstm, above=of g3] (b3) {};
	\node[lstm, above=of g4] (b4) {};
	\draw [chainLine] (b4) to (b3);
	\draw [chainLine] (b3) to (b2);
	\draw [chainLine] (b2) to (b1);
	\draw [chainLine] (x1) to [out=125,in=-125] (b1);
	\draw [chainLine] (x2) to [out=125,in=-125] (b2);	
	\draw [chainLine] (x3) to [out=125,in=-125] (b3);
	\draw [chainLine] (x4) to [out=125,in=-125] (b4);
	\node[hidden, above= of b1] (h1) {${\rm h}^{fb}_{i,1}$};
	\node[hidden, above= of b2] (h2) {${\rm h}^{fb}_{i,2}$};
	\node[hidden, above= of b3] (h3) {${\rm h}^{fb}_{i,3}$};
	\node[hidden, above= of b4] (h4) {${\rm h}^{fb}_{i,4}$};
	\draw [chainLine] (b1) to node [left, text=black] {${\rm h}_{i,1}^b$} (h1);
	\draw [chainLine] (b2) to node [left, text=black] {${\rm h}_{i,2}^b$} (h2);
	\draw [chainLine] (b3) to node [left, text=black] {${\rm h}_{i,3}^b$} (h3);
	\draw [chainLine] (b4) to node [left, text=black] {${\rm h}_{i,4}^b$} (h4);
	\draw [chainLine] (g1) to [out=35,in=-35] node [left, text=black, yshift=12mm, xshift=4mm] {${\rm h}_{i,1}^f$} (h1);
	\draw [chainLine] (g2) to [out=35,in=-35] node [left, text=black, yshift=12mm, xshift=4mm] {${\rm h}_{i,2}^f$} (h2);
	\draw [chainLine] (g3) to [out=35,in=-35] node [left, text=black, yshift=12mm, xshift=4mm] {${\rm h}_{i,3}^f$} (h3);
	\draw [chainLine] (g4) to [out=35,in=-35] node [left, text=black, yshift=12mm, xshift=4mm] {${\rm h}_{i,4}^f$} (h4);
	\node[mlp, above=of h1] (m1) {};
	\node[mlp, above=of h2] (m2) {};
	\node[mlp, above=of h3] (m3) {};
	\node[mlp, above=of h4] (m4) {};
	\draw [chainLine] (h1) to (m1);
	\draw [chainLine] (h2) to (m2);
	\draw [chainLine] (h3) to (m3);
	\draw [chainLine] (h4) to (m4);
	\node[textNode, above=of m1] (y1) {$\bar{y}_{i,1}$};
	\node[textNode, above=of m2] (y2) {$\bar{y}_{i,2}$};
	\node[textNode, above=of m3] (y3) {$\bar{y}_{i,3}$};
	\node[textNode, above=of m4] (y4) {$\bar{y}_{i,4}$};
	\draw [chainLine] (m1) to (y1);
	\draw [chainLine] (m2) to (y2);
	\draw [chainLine] (m3) to (y3);
	\draw [chainLine] (m4) to (y4);
	\end{tikzpicture}
	}
	\caption{Bidirectional LSTM neural network architecture. The model is composed of two rows of LSTM cells with information propagating from the left in the bottom row and from the right in the top row. The information from the LSTM cells is concatenated in ${\rm h}_{i,l}^{fb}$ and then processed in the multi-layer perceptrons MLP to return the predicted output $\bar{y}_{i,l}$. }
	\label{fig:model}
\end{figure}

Recurrent neural networks (RNNs) are a family of neural networks for sequence-to-sequence task. 
Different from CRF, it takes a sequence of vectors $\mathbf{X}_i = \left(\vec{\rm x}_{i,1}, \vec{\rm x}_{i,2}, \cdots, \vec{\rm x}_{i,L}\right)$ as input and return a sequence of hidden output vectors $\mathbf{H} = \left(\vec{\rm h}_{i,1} , \vec{\rm h}_{i,2}, \cdots, \vec{\rm h}_{i,L}\right)$. 
In the case discussed in Sec. II.B of the main paper, a sequence $\vec{x}_i$ becomes a sequence of vectors $\mathbf{X}_i = \left( \left(x_{i,1}, \sqrt{1-x_{i,1}^2}\right), \cdots, \left(x_{i,L}, \sqrt{1-x_{i,L}^2}\right)\right)$. 
Theoretically, traditional RNNs can learn long dependencies as the output at each position depends on information from previous positions, although an effective implementation may be difficult~\cite{bengio1994learning}. 
Long short-term memory networks~\cite{HochreiterSchmidhuber1997} are designed to provide an effective solution by using a memory-cell and they have been shown to capture long-range dependencies. 
Several gates are used to control the proportion of the information sent to the memory cell and proportion of information to forget. 
Specifically, for a one directional LSTM cell we use the implementation 
\begin{align}
\vec{\rm i}_{i,l} &= \sigma \left(\mathbf{W}_{\rm xi}\; \vec{\rm x}_{i,l} + \vec{\rm b}_{\rm xi} + \mathbf{W}_{\rm hi}\; \vec{\rm h}_{i,l-1} + \vec{\rm b}_{\rm hi}\right) \\
\vec{\rm f}_{i,l} & = \sigma \big(\mathbf{W}_{\rm xf}\; \vec{\rm x}_{i,l} + \vec{\rm b}_{\rm xf} + \mathbf{W}_{\rm hf}\; \vec{\rm h}_{i,l-1} + \vec{\rm b}_{\rm hf}\big) \\
\vec{\rm g}_{i,l} & = \tanh \big(\mathbf{W}_{\rm xg}\; \vec{\rm x}_{i,l} + \vec{\rm b}_{\rm xg} + \mathbf{W}_{\rm hg}\; \vec{\rm h}_{i,l-1} + \vec{\rm b}_{\rm hg}\big) \\
\vec{\rm o}_{i,l} & = \sigma  \big(\mathbf{W}_{\rm xo}\; \vec{\rm x}_{i,l} + \vec{\rm b}_{\rm xo} + \mathbf{W}_{\rm ho}\; \vec{\rm h}_{i,l-1} + \vec{\rm b}_{\rm ho}\big) \\
\vec{\rm c}_{i,l} & = \vec{\rm f}_{i,l} \odot \vec{\rm c}_{i,l-1} + \vec{\rm i}_{i,l} \odot \vec{\rm g}_{i,l}\\
\vec{\rm h}_{i,l} & = \vec{\rm o}_{i,l} \odot \tanh(\vec{\rm c}_{i,l})
 \end{align}
where $\sigma$ is the element-wise sigmoid function and $\odot$ is the element-wise product. Each $\mathbf{W}$ matrix represents the parameters of the LSTM. The $\mathbf{W}_{\rm xa}$ are matrices with an hidden size $d_{LSTM}$ number of rows and two columns, the $\mathbf{W}_{\rm xa}$ have $d_{LSTM}$ rows and columns. The $\vec{b}_{\rm xa}$ and $\vec{b}_{\rm xa}$ are matrices with $d_{LSTM}$ rows and $d_B$ columns. Here the ``${\rm a}$'' can be one of the labels ${\rm i,\;f,\;g}$ or ${\rm o}$. 
$\vec{\rm i}_{i,l}, \vec{\rm f}_{i,l}, \vec{\rm g}_{i,l} $ and $\vec{\rm o}_{i,l}$ are, respectively, the input, forget, cell and output gates at position $l$. 
 
However, in order to capture better the correlations both to the right and to the left of a certain position $l$, we use a bidirectional LSTM~\cite{graves2005framewise}, with forward and backward LSTM pairs of cells as shown in Fig.\ref{fig:model}.  
 
The representation of a specific position $l$ is given by the concatenation of the forward and backward LSTM, respectively each providing hidden output vectors $\vec{h}_{i,l}^f$ and $\vec{h}_{i,l}^b$, to give the overall forward-backward hidden output vector  $\vec{h}_{i,l}^{fb} = \big[ \vec{h}_{i,l}^f;  \vec{h}_{i,l}^b\big]$. 
 
In order to obtain a scalar at each time step, we use a final non-linear layer (multi-layer percerptron MLP) to map the hidden vector to a scalar value at each position $l$  
\begin{equation}
\bar{y}_{i,l} = \sigma \big(\vec{w}^{\mathbf{T}}_{\rm hy} \; \vec{\rm h}_{i,l}^{fb} + b_{\rm hy}\big). 
\end{equation}

Once the sequence of $\bar{y}_{i,l}$ is generated, the optimization of the model parameters is achieved by minimizing the error between predicted $\vec{\bar{y}}_i$ and exact sequences $\vec{y}_i$.

\end{appendix} 


\begin{thebibliography}{99} 


\bibitem{SchuldPetruccione2014} M. Schuld, I. Sinayskiy, and F. Petruccione, Contemporary Physics {\bf 56}, 172 (2014). 
\bibitem{AdcockStanisic2015} J. Adcock, E. Allen, M. Day, S. Frick, J. Hinchliff, M. Johnson, S. Morley-Short, S. Pallister, A. Price, and S. Stanisic, arXiv:1512.02900 (2015). 
\bibitem{BiamonteLloyd2016} J. Biamonte, P. Wittek, N. Pancotti, P. Rebentrost, N. Wiebe, and S. Lloyd, Nature {\bf 549}, 195 (2017).
\bibitem{KalininArchibald2015} S. V. Kalinin, B. G. Sumpter, and R. K. Archibald, Nat. Mater. {\bf 14}, 973 (2015). 
\bibitem{SchoenholzLiu2016} S.S. Schoenholz, E.D. Cubuk, D.M. Sussman, E. Kaxiras, and A.J. Liu, Nature Phys {\bf 12}, 469 (2016). 
\bibitem{Wang2016} L. Wang, Phys. Rev. B {\bf 94}, 195105 (2016). 
\bibitem{CarrasquillaMelko2016} J. Carrasquilla, and R.G. Melko, Nature Physics {\bf 13}, 431 (2017).  
\bibitem{ZhangKim2017} Y. Zhang, and E.A. Kim, Phys. Rev. Lett. {\bf 118}, 216401 (2017).     
\bibitem{ChngKhatami2018} K. Ch'’ng, N. Vazquez, and E. Khatami, Phys. Rev. E {\bf 97}, 013306 (2018). 
\bibitem{ZhangZhai2018} P. Zhang, H. Shen, and H. Zhai, Phys. Rev. Lett. {\bf 120}, 066401 (2018).       
\bibitem{NieuwenburgHuber2017} E. P. van Nieuwenburg, Y.-H. Liu, and S. D. Huber, Nat. Phys. {\bf 13}, 435 (2017).   
\bibitem{BroeckerTrebst2016} P. Broecker, J. Carrasquilla, R. G. Melko, and S. Trebst, Scientific Reports {\bf 7}, 8823 (2017).      

\bibitem{ChngKhatami2016} K. Ch’'ng, J. Carrasquilla, R. G. Melko, and E. Khatami, Phys. Rev. X {\bf 7}, 031038 (2017).
\bibitem{ArsenaultMillis2014} L.F. Arsenault, A. Lopez-Bezanilla, O.A. von Lilienfeld, and A.J. Millis, Phys. Rev. B {\bf 90}, 155136 (2014). 
\bibitem{ArsenaultMillis2015} L.-F. Arsenault, O. A. von Lilienfeld, and A. J. Millis, arXiv:1506.08858 (2015).
\bibitem{TorlaMelko2016} G. Torlai and R. G. Melko, Phys. Rev. B 94, 165134 (2016).
\bibitem{AminMelko2016} M. H. Amin, E. Andriyash, J. Rolfe, B. Kulchytskyy, and R. Melko, arXiv:1601.02036.    
\bibitem{LiuPu2016} J. Liu, Y. Qi, Z. Y. Meng, and L. Fu, Phys. Rev. B {\bf 95}, 041101 (2017).
\bibitem{HuangWang2016} L. Huang and L. Wang, Phys. Rev. B {\bf 95}, 035105 (2017).
\bibitem{AokiKobayashi2016} K.-I. Aoki and T. Kobayashi, Mod. Phys. Lett. B, 1650401 (2016). 
\bibitem{CarleoTroyer2017} G. Carleo, and M. Troyer, Science {\bf 355}, 602 (2017). 
\bibitem{NomuraImada2017} Y. Nomura, A.S. Darmawan, Y. Yamaji, and M. Imada, Phys. Rev. B {\bf 96}, 205152 (2017). 
\bibitem{CzischekGasenzer2018} S. Czischek, M. G\"rttner, and T. Gasenzer, arxiv:1803.08321.  

\bibitem{Beny2013} C. Beny, arXiv:1301.3124. 
\bibitem{MethaSchwab2014} P. Mehta and D. J. Schwab, arXiv:1410.3831. 
\bibitem{LinRolnick2017} H.W. Lin, M. Tegmark, and D. Rolnick, Journal of Statistical Physics {\bf 168}, 1223 (2017). 



\bibitem{StoudenmireSchwab2016} E.M. Stoudenmire and D.J. Schwab, Advances In Neural Information Processing Systems {\bf 29}, 4799 (2016).         
\bibitem{HanZhang2017} Z.-Y. Han, J. Wang, H. Fan, L. Wang, and P. Zhang, arxiv:1709.01662. 
\bibitem{Stoudenmire2017} E.M. Stoudenmire, arXiv:1801.00315 (2017). 
\bibitem{NovikovOseledets2017} A. Novikov, M. Trofimov, and I. Oseledets, arxiv:1605.03795, ICLR (2017). 
\bibitem{YangGu2018} C. Yang, F.C. Binder, V. Narasimhachar, M. Gu, arxiv:1803.08220.         
\bibitem{PestunVlassopoulos2017} V. Pestun, J. Terilla, and Y. Vlassopoulos, arxiv:1711.01416.  
\bibitem{DengSarma2017} D.L. Deng, X. Li, and S.D. Sarma, Physical Review X {\bf 7}, 021021 (2017).  
\bibitem{ChenXiang2018} J. Chen, S. Cheng, H. Xie, L. Wang, and T. Xiang, Phys. Rev. B {\bf 97}, 085104 (2018).  


\bibitem{Cichocki2014} A. Cichocki, arxiv:1403.2048.      
\bibitem{Schollwock2011} U. Schollw\"ock, Annals of Physics {\bf 326}, 96 (2011). 
\bibitem{DerridaPasquier1993} B. Derrida, M.R. Evans, V. Hakim and V. Pasquier, J. Phys. A: Math. Gen. {\bf 26}, 1493 (1993). 
\bibitem{KrebsSandov1997} K. Krebs and S. Sandow, J. Phys. A: Math. Gen. {\bf 30}, 3165 (1997).    
\bibitem{JohnsonJaksch2010} T.H. Johnson, S.R. Clark, and D. Jaksch, Phys. Rev. E {\bf 82}, 036702 (2010). 
\bibitem{JohnsonJaksch2015} T.H. Johnson, T.J. Elliott, S.R. Clark, and D. Jaksch, Phys. Rev. Lett. {\bf 114}, 090602 (2015). 



\bibitem{White1992} S.R. White, Phys. Rev. Lett. {\bf 69}, 2863 (1992). 
\bibitem{Schollwock2005} U. Schollw\"ock, Rev. Mod. Phys. {\bf 77}, 259 (2005). 
\bibitem{McCulloch2007} I.P. McCulloch, J. Stat. Mech., P10014 (2007). 
\bibitem{WhiteFeguin2004} S.R. White and A.E. Feiguin, Phys. Rev. Lett. {\bf 93}, 076401 (2004).  
\bibitem{Vidal2004} G. Vidal, Phys. Rev. Lett. {\bf 93}, 040502 (2004).  
\bibitem{DaleyVidal2004} A.J. Daley, C. Kollath, U. Schollw\"ock, and G. Vidal, J. Stat. Mech.: Theor. Exp. P04005 (2004).    
\bibitem{VerstraeteCirac2004} F. Verstraete, J.J. Garcia-Ripoll, and J.I. Cirac, Phys. Rev. Lett. {\bf 93}, 207204 (2004).    
\bibitem{Daley2014} A.J. Daley, Adv. Phys. {\bf 63}, 77 (2014).  


\bibitem{LaffertyPereira2001} J. Lafferty, A. McCallum, and F. Pereira, Proc. 18th International Conf. on Machine Learning. Morgan Kaufmann. 282 (2001). 
\bibitem{HochreiterSchmidhuber1997} S. Hochreiter and J. Schmidhuber, Neural Computation {\bf 9}, 1735 (1997).


\bibitem{Wolfram1983} S. Wolfram, Rev. Mod. Phys. {\bf 55}, 601 (1983).        
\bibitem{fn1} If $l+j>L$ then the value does not change. 
\bibitem{ProsenBuca2017} T. Prosen, and B. Buca, J. Phys. A: Math. Theor. {\bf 50}, 395002 (2017). 
\bibitem{fn4} Note also that matrix product operators for unitary evolutions are quantum cellular automata \cite{CiracVerstraete2017}. 
\bibitem{GersCummins2000} F.A. Gers, J. Schmidhuber, and F. Cummins, Neural Computation {\bf 12}, 2451 (2000).    
\bibitem{graves2005framewise} A. Graves and J. Schmidhuber, IEEE International Joint Conference on Neural Networks Proceedings, 2047 (2005). 
\bibitem{kingba2014} D.P. Kingma, and J. Ba, arXiv:1412.6980 (2014) 


\bibitem{fn2} In our calculations the sites with the $1$ where sites $l=1+78\delta$ where $\delta$ is one of the possible digits $0,\;1,\dots, 9$. The choice of $78$ is due to the size of the images used which have $28\times 28=784$ pixels, which are then divided by $10$ which is the number of possible digits in the classification. 
\bibitem{fn3} More precise results could be obtained by optimizing the location of the ones. 
\bibitem{ChristopherCortes} J.C. Christopher, Y.L. Burges, C. Cortes, “MNIST handwritten digit database”, http://yann.lecun.com/exdb/mnist/ (1998). 

\bibitem{CiracVerstraete2017} J.I. Cirac, D. Perez-Garcia, N. Schuch, F. Verstraete, J. Stat. Mech. 083105 (2017).  





\end{thebibliography}

\begin{thebibliography}{99} 

\bibitem[A1]{ma2016end} X. Ma and E. Hovy, Proceedings of ACL, 1064 (2016).  
\bibitem[A2]{liu1989limited} D. Liu and J. Nocedal, Mathematical programming  {\bf 45}, 503 (1989). 
\bibitem[A3]{bengio1994learning} Y. Bengio, P. Simard and P. Frasconi, IEEE transactions on neural networks  {\bf 5}, 157 (1994). 

\end{thebibliography}
\end{document}